\documentclass[amsmath,amssymb,aps,prd,tightenlines,superscriptaddress,
nofootinbib,preprintnumbers,notitlepage]{revtex4-1}
\usepackage{multirow}
\usepackage{graphicx}
\usepackage{tabularx}
\usepackage{slashed}
\usepackage{url}
\usepackage{hyperref}
\usepackage{rotating}
\usepackage{braket}
\usepackage{color,xcolor}
\usepackage{booktabs}
\usepackage[normalem]{ulem}

\newcommand{\wilson}{c}

\newcommand{\ord}{\ensuremath{\mathcal{O}}}

\newcommand{\lag}{\ensuremath{\mathcal{L}}}

\newcommand{\sw}{\ensuremath{s_w}}
\newcommand{\swd}{\ensuremath{s^2_w}}
\newcommand{\cw}{\ensuremath{c_w}}
\newcommand{\cwd}{\ensuremath{c^2_w}}

\newcommand{\sad}{\ensuremath{s^2_\alpha}}
\newcommand{\sa}{\ensuremath{s_\alpha}}

\newcommand{\aem}{\ensuremath{\alpha_{\text{em}}}}
\newcommand{\aew}{\ensuremath{\alpha_{\text{ew}}}}

\newcommand{\mwd}{\ensuremath{m_W^2}}
\newcommand{\mzd}{\ensuremath{m_Z^2}}

\newcommand{\qqquad}{\qquad \qquad}
\newcommand{\qqqquad}{\qquad \qquad \qquad}

\newcommand{\st}[1]{\tilde{t}_{#1}}

\newcommand{\msb}[1]{m_{\tilde{b}_{#1}}}
\newcommand{\mst}[1]{m_{\tilde{t}_{#1}}}

\def\slashchar#1{\setbox0=\hbox{$#1$}           % set a box for #1
   \dimen0=\wd0                                 % and get its size
   \setbox1=\hbox{/} \dimen1=\wd1               % get size of /
   \ifdim\dimen0>\dimen1                        % #1 is bigger
      \rlap{\hbox to \dimen0{\hfil/\hfil}}      % so center / in box
      #1                                        % and print #1
   \else                                        % / is bigger
      \rlap{\hbox to \dimen1{\hfil$#1$\hfil}}   % so center #1
      /                                         % and print /
   \fi}

\def\eg{\textit{e.g. }}
\def\ie{\textit{i.e. }}
\newcommand{\obw}{\ensuremath{\mathcal{O}}_{B,W}}
\newcommand{\oh}{\ensuremath{\mathcal{O}}_H}

\newcommand{\oT}{\ensuremath{\mathcal{O}}_T}

\newcommand{\pbp}{\ensuremath{\phi^\dagger\,\phi}}

\newcommand{\lageff}{\ensuremath{\lag_{\text{eff}}}}

\newcommand{\stone}{{\ensuremath{\tilde{t}_{1}}}}
\newcommand{\sttwo}{{\ensuremath{\tilde{t}_{2}}}}

\newcommand{\Qtilde}{\ensuremath{\tilde{Q}}}
\newcommand{\TR}{\ensuremath{\tilde{t}_R}}

\newcommand{\sbR}{\ensuremath{\tilde{b}_{R}}}

\newcommand{\kLL}{\ensuremath{\kappa_{LL}}}
\newcommand{\kLR}{\ensuremath{\kappa_{LR}}}
\newcommand{\kRR}{\ensuremath{\kappa_{RR}}}

\newcommand{\kLLtilde}{\ensuremath{\tilde{\kappa}_{LL}}}
\newcommand{\kRRtilde}{\ensuremath{\tilde{\kappa}_{RR}}}

\newcommand{\mhd}{\ensuremath{m_h^2}}
\newcommand{\mh}{\ensuremath{m_h}}
\newcommand{\mHHd}{\ensuremath{m_H^2}}
\newcommand{\mHH}{\ensuremath{m_H}}

\newcommand{\ctd}{\ensuremath{c^2_{\tilde{t}}}}
\newcommand{\std}{\ensuremath{s^2_{\tilde{t}}}}
\newcommand{\sdt}{\ensuremath{s_{2 \tilde{t}}}}

\newcommand{\ct}{\ensuremath{c_{\tilde{t}}}}
\renewcommand{\st}{\ensuremath{s_{\tilde{t}}}}

\newcommand{\cft}{\ensuremath{c_{4 \tilde{t}}}}

\newcommand{\ceit}{\ensuremath{c_{8 \tilde{t}}}}

\newcommand{\mstone}{\ensuremath{m_{\tilde{t}_1}}}
\newcommand{\msttwo}{\ensuremath{m_{\tilde{t}_2}}}

\newcommand{\uvpole}{\ensuremath{\Delta_\epsilon}}

\newcommand{\MLeft}{\ensuremath{M_{\tilde{Q}_L}}}
\newcommand{\MRight}{\ensuremath{M_{\tilde{T}_R}}}
\newcommand{\epsaa}{\ensuremath{\epsilon_{\gamma\gamma}}}
\newcommand{\PHiggs}{\ensuremath{H}}
\newcommand{\Azero}{\ensuremath{A^0}}

\AtBeginDocument{
  \heavyrulewidth=.08em
  \lightrulewidth=.05em
  \cmidrulewidth=.03em
  \belowrulesep=.65ex
  \belowbottomsep=0pt
  \aboverulesep=.4ex
  \abovetopsep=0pt
  \cmidrulesep=\doublerulesep
  \cmidrulekern=.5em
  \defaultaddspace= .5em
  \setlength{\tabcolsep}{0.5em}
}

\begin{document}

%%%%%%%%%%%%%%%%%%%%%%%%%%%%%%%%%%%%%%%%%%%%%%%%%%%%%%%%%%%%
%\title{Improved Matching for Extra Scalars in Effective Lagrangians} 
\title{Matching Matters!}

\author{Ayres Freitas}
\affiliation{PITT-PACC, Department of Physics \& Astronomy, University of Pittsburgh, USA}

\author{David L\'opez-Val}
\affiliation{Centre for Cosmology, Particle Physics \& Phenomenology CP3, Universit\'e catholique de Louvain, Belgium}
 
\author{Tilman Plehn}
\affiliation{Institut f\"ur Theoretische Physik, Universit\"at Heidelberg, Germany}
\date{\today}

\begin{abstract}  
Effective Lagrangians are a useful tool for a data-driven approach to
physics beyond the Standard Model at the LHC. However, for the new
physics scales accessible at the LHC, the effective operator expansion
is only relatively slowly converging at best. For tree-level
processes, it has been found that the agreement between the effective
Lagrangian and a range of UV-complete models depends sensitively on
the appropriate definition of the matching.  We extend this analysis
to the one-loop level, which is relevant for electroweak precision
data and Higgs decay to photons. 
We show that near the scale of
electroweak symmetry breaking the validity of the effective theory description
can be systematically improved through an appropriate matching
procedure.
In particular, we find a significant increase in accuracy when
including suitable terms suppressed by the Higgs vacuum expectation value in
the matching. 
\end{abstract}

\maketitle
\tableofcontents

\clearpage

%%%%%%%%%%%%%%%%%%%%%%%%%%%%%%%%%%%%%%%%%%%%%%%%%%%%%%%%%%%%
\section{Introduction}
\label{sec:intro}
%%%%%%%%%%%%%%%%%%%%%%%%%%%%%%%%%%%%%%%%%%%%%%%%%%%%%%%%%%%%

After the end of the LHC Run~I and with the start of Run~II, 
the field of particle physics has clearly entered a
data-driven era. While we should not entirely dismiss our theoretical
or experimental motivations to search for specific models of physics
beyond the Standard Model, the amount of available LHC data requires a
more model-independent language to analyze and communicate
experimental results. This has lead the
Higgs~\cite{higgs_eft,silh,legacy},
electroweak~\cite{ew_eft,Hagiwara:1993ck}, top~\cite{top_eft}, and
dark matter~\cite{dm_eft} communities to employ effective Lagrangians
or related methods to communicate LHC results.  Another good example
is the recent hint for a 750~GeV resonance, where the limited
available data at best allows for an effective theory
analysis. Nevertheless, the theory community also illustrated the
limits of the effective theory approach by immediately linking the LHC
anomaly to any number of models.

This strategy implies that independent of the effective theory being
the main theoretical description of a given physics sector, the
effective Lagrangian can serve as a means of communication between
experiment and theory. To this end it is not necessary to show that an
effective theory at the LHC is a fully consistent theory framework;
instead, a given effective Lagrangian has to describe the effects of
classes of new physics models at the LHC within the expected
experimental precision.  For Higgs signatures with a wide range of
kinematic configurations this question has been studied at length,
both for strongly and weakly interacting
models~\cite{Biekoetter:2014jwa,heft_limitations,heft_limitations2,Brehmer:2015rna,Biekotter:2016ecg,Drozd:2015kva}. It
turns out that the expected measurement uncertainties largely limit us
to tree-level effects of new physics, and that an appropriately
defined dimension-6 Lagrangian description only breaks down in the
presence of new resonance features.\bigskip

In this paper we extend our effective Lagrangian considerations to
systematically include quantum effects. We start by introducing two
ways of improving the matching of the effective Lagrangian in
Sec.~\ref{sec:vimproved} and Sec.~\ref{sec:brokenphase}. Both of them
target the problem that in the relevant region of parameter space the
effective Lagrangian does not have a large scale hierarchy and instead
we have to work under the weak condition that after electroweak
symmetry breaking the new particles lie just above the weak scale, $v
\lesssim \Lambda$.

We then study heavy particle loops contributing to electroweak
precision observables (EWPO) in Sec.~\ref{sec:oblique}. Two
representative models for extended scalar sectors allow us to study
the underlying features: an additional scalar electroweak singlet, and
color-triplet heavy-quark scalar partners.  We compute the oblique
electroweak precision observables $S$ and $T$ in the full, UV-complete
model as well as based on the effective Lagrangian. For the latter we
explore several prescriptions for the one-loop matching.  With the
Higgs portal, contributions from loop-induced operators combine with
loop insertions of tree-level operators. For the top partners all new
physics effects appear through virtual heavy scalars and loop-induced
operators.  In addition, top partners feature in general two heavy scales,
allowing us to test a dimension-6 description in the presence of
multiple mass scales. All these are challenges which our matching
prescription for the effective Lagrangian has to face.

Secondly, we study the loop-induced Higgs coupling to photons in
Sec~\ref{sec:photons}. Because a singlet Higgs portal hardly shows any
features in this observable, we now test a two-Higgs doublet model
including a charged Higgs boson, as well as the scalar top partner model
mentioned above. Again, we show how the
choice of matching procedure can significantly and systematically
improve the agreement between the effective Lagrangian and the full
models.

%%%%%%%%%%%%%%%%%%%%%%%%%%%%%%%%%%%%%%%%%%%%%%%%%%%%%%%%%%%%
\subsection{Effective Lagrangian}
\label{sec:heft}

Effective field theories provide a useful language to communicate
experimental results without having to specify any details of an
underlying model. At energies below the characteristic UV scale of the
new physics sector, only the light states are the physically
accessible degrees of freedom. Based on the dynamic degrees of freedom
at low energies, symbolically denoted as $\phi$, and the underlying
symmetries we define a Lagrangian of the kind
\begin{align}
  \lageff &= \lag_\text{SM} + \sum_{d=5}^\infty \, \sum_{a_d} \,
            \dfrac{\wilson_{a_d}^{(d)}}{\Lambda^{d-4}}\,\mathcal{O}_{a_d}^{(d)} \; ,
  \label{eq:efflaggen}
\end{align}
where the heavy field dynamics is described by the Wilson coefficients
$\wilson_i(\mu)$. The effective operators $\mathcal{O}_i(\mu)$
parametrize the local interactions among the light states. The
effective Lagrangian of Eq.\eqref{eq:efflaggen} follows from averaging
over short distance effects, which in the functional formalism of QFT
means integrating out of the heavy field fluctuations in the UV action
path integral~\cite{functional,hlm}.

To relate the Wilson coefficients $\wilson_i(\mu)$ to a set of full
model parameters $g_j(\mu)$ we use the fact that, by construction, the
effective Lagrangian reproduces the full model predictions in the
low-energy range $E < \Lambda$. In quantum field theory observables
are derived from one-particle-irreducible (1PI) $n$-point Greens
functions. Therefore, we compute a set of renormalized 1PI Greens
functions in the full model and based on the effective Lagrangian
setups with help of the packages \textsc{FeynArts} and
\textsc{FormCalc}~\cite{feynarts}.  Both of them we evaluate at an
appropriate matching scale $\mu = \Lambda$,
\begin{align}
\Gamma^\text{1PI}_\text{full}\,  [\phi]\,( g_j, \mu=\Lambda) = 
\Gamma^\text{1PI}_\text{EFT}\,[\phi]\,( \wilson_i, \mu=\Lambda) \; .
\label{eq:matching-generic}
\end{align}
The matching scale $\Lambda$ is usually identified with the
characteristic UV scale of the effective Lagrangian, above which the
high-energy degrees of freedom start to be resolved.\footnote{For LHC
  processes this statement is signature-dependent, because 
  particles appearing, say, in the $s$-channel are much easier to resolve
  than particles in the $t$-channel. Similarly, particles appearing in
  loops are much harder to resolve than particles appearing at tree
  level. Clearly, any statement considering the quantitative validity
  of an effective Lagrangian needs to take these differences into
  account~\cite{Brehmer:2015rna}.}
Equation~\eqref{eq:matching-generic} allows us to express each Wilson
coefficient in terms of the model parameters.  For weakly coupled
theories, the matching condition is applied order-by-order in the
perturbative expansion, identifying the tree-level 1PI graphs first,
and then moving to one-loop and beyond.\bigskip

In this paper we assume a linear realization of electroweak symmetry
breaking with a vacuum expectation value (vev) $v=246$~GeV. We truncate our
set of operators at dimension-6, which has been shown to be sufficient
to describe (most of) the expected LHC observables. Some popular
bases of these dimension-6 operators are the
Warsaw~\cite{Grzadkowski:2010es}, HISZ~\cite{Hagiwara:1993ck}, and
SILH bases~\cite{silh}. All three maximize the use of bosonic
operators to describe Higgs and electroweak observables. They can be
mapped onto each other using equations of motion, integration by
parts, field redefinitions, and Fierz
transformations~\cite{Alonso:2014rga}.  We use the SILH basis and
retain only those operators relevant for Higgs physics at the
LHC~\cite{silh}.  The effective Lagrangian truncated to dimension 6
reads
\begin{align}
  \lag_\text{EFT} = \lag_\text{SM} 
&+ \frac{\wilson_H}{2\Lambda^2} \, \partial^\mu (\pbp) \, \partial_\mu (\pbp)
+ \frac{\wilson_T}{2\Lambda^2} \, (\phi^\dagger \, \overleftrightarrow{D}^\mu \, \phi) \, (\phi^\dagger \, \overleftrightarrow{D}_\mu \, \phi)
-\frac{\wilson_6\lambda}{\Lambda^2} (\pbp)^3 \notag \\
&+ \frac{ig \, \wilson_W}{2 \Lambda^2} \, (\phi^\dagger \, \sigma^k \overleftrightarrow{D}^\mu\phi) \,  D^\nu \, {W^k}_{\mu\nu}
                         + \frac{ig'\wilson_B}{2\Lambda^2} \, (\phi^\dagger \, \overleftrightarrow{D}^\mu \, \phi) \, \partial^\nu \, B_{\mu\nu} \notag \\
&+ \frac{ig \, \wilson_{HW}}{\Lambda^2} \, (D^\mu \, \phi^\dagger) \, \sigma^k \, (D^\nu \, \phi) \, W^k_{\mu\nu}
                       + \frac{ig'\wilson_{HB}}{\Lambda^2} (D^\mu\phi^\dagger) \, (D^\nu \, \phi) \, B_{\mu\nu}\notag \\
&+ \frac{g'^2 \wilson_\gamma}{\Lambda^2} \, (\pbp) \, B_{\mu\nu} \, B^{\mu\nu} + \frac{g_s^2 \wilson_g}{\Lambda^2} \, (\pbp) \, G^A_{\mu\nu} \, G^{\mu\nu\, A} \notag \\
&- \left[
  \frac{\wilson_u}{\Lambda^2} \, y_u \, (\pbp) (\phi^\dagger\cdot \, \overline{Q}_L) \, u_R
+ \frac{\wilson_d}{\Lambda^2} \, y_d \, (\pbp) (\phi \,  \overline{Q}_L) \, d_R  
+ \frac{\wilson_\ell}{\Lambda^2} \, y_\ell \, (\pbp) (\phi \,  \overline{L}_L) \, \ell_R
                                 + \text{h.c.} \, \right] \,.
\label{eq:EFT}
\end{align}
Here, $g = e/\sw, g' = e/\cw$, and $g_s$ stand for the SM gauge
couplings and $\lambda$ denotes the usual Higgs quartic. The
dimension-6 Wilson coefficients $\wilson_i$ are defined with a
universal suppression of $1/\Lambda^2$ rather than the ad-hoc prior of
$1/v^2$ or $1/m_W^2$ in the original proposal. The notation of the
individual operators follows the notation for the Wilson coefficients,
for example $\oT \sim (\phi^\dagger \, \overleftrightarrow{D}^\mu \,
\phi) \, (\phi^\dagger \, \overleftrightarrow{D}_\mu \, \phi)$.

%%%%%%%%%%%%%%%%%%%%%%%%%%%%%%%%%%%%%%%%%%%%%%%%%%%%%%%%%%%%
\subsection{$v$-improved matching}
\label{sec:vimproved}

Whenever we discuss the validity of effective Lagrangian approaches we
need to keep in mind that the matching of the individual Wilson
coefficients to a given full model is not defined uniquely. This is
particularly true when the matching scale is not far from the scale at
which the electroweak symmetry is broken. A hierarchy of scales
$\Lambda \gg v$ certainly justifies that we define an effective action
by integrating out all heavy states in the unbroken phase $\langle
\phi \rangle = 0$ and truncate that action only including terms up to
the order $1/\Lambda^2$. The Wilson coefficients are by construction
independent on the light field masses and low-scale parameters. In
this ideal world of new physics governed by a single, very large
energy scale, the default matching is free from ambiguities and leads
to a rapidly converging effective field theory.

However, at hadron colliders, the numerator compensating the inverse
powers of $\Lambda$ can be any parameter with the appropriate mass
dimension, including the partonic collider energy or masses induced by
electroweak symmetry breaking. In addition, the relevant heavy mass
scale does not have to be a Lagrangian parameter in some interaction
basis, it can also be particle masses induced by other mechanism than
electroweak symmetry breaking. Examples are vector fermion masses or
supersymmetric partner masses, which receive dominant contributions
from some heavy Lagrangian parameter, but subdominant effects from the
electroweak vev. In this case, electroweak symmetry breaking will in
general induce additional scales $\Lambda \pm gv$ from the new physics
couplings to the Higgs. The default operator expansion in the unbroken
phases removes all contributions of the type $v/\Lambda$ from the
definition of the matching scale and from the dimension-6 Wilson
coefficients.  If these corrections should be non-negligible it will
fail to capture features of the full model. In that case, we can adapt
the details of the matching scheme to enhance the level of agreement
between the effective Lagrangian and the full
model~\cite{Brehmer:2015rna}.

In our tree level analysis, we account for $v$-induced effects in two
ways~\cite{Brehmer:2015rna}: first, the matching scale $\Lambda$ is
not identified with a Lagrangian mass parameter in the unbroken phase,
but with the mass of a physical particle.  The masses of the heavy
states can lead to more than one heavy scale, with the splitting
generated by the electroweak vev. In this case, $\Lambda$ is
identified with the lightest new state by default.

Second, we allow for corrections suppressed by $v/\Lambda$ in the
Wilson coefficients.  On the full model side we express the relevant
observables in terms of model parameters in the mass-eigenstate
Lagrangian, \ie in terms of masses and mixing angles of the physical
states. In this way, one effectively includes corrections from some
higher-dimensional operators of the form $\mathcal{O}^{d=6+n} \sim
\mathcal{O}^{d=6}\,\times\,(\phi^\dagger\phi)^n$, where the Higgs
doublets appear as $v$-insertions in the broken phase.\bigskip

We use the name $v$-improvement for the combination of these two steps.
It is worth emphasizing that the $v$-improved prescription does not
introduce additional free parameters, nor does it break any symmetries
of the original Lagrangian. In that sense, it describes an equivalent
effective Lagrangian. The $v$-improved matching of a
linear realization of electroweak symmetry breaking is also different
from the non-linear realization, because we still require that the
Higgs boson forms a weakly coupled doublet with the Goldstone bosons.

%%%%%%%%%%%%%%%%%%%%%%%%%%%%%%%%%%%%%%%%%%%%%%%%%%%%%%%%%%%%
\subsection{Broken-phase matching}
\label{sec:brokenphase}

While the $v$-improvement described above is sufficient to
systematically improve the agreement between the full model and the
dimension-6 approximation, additional complications arise for
loop-induced processes. As mentioned above, the functional approach to
the effective Lagrangian is straightforward if the heavy sector can be
fully separated from the light fields and integrated out in the path
integral.  In the case of mass eigenstates affected by mixing of heavy
and a light field components, diagrammatic methods can improve the
matching between the full theory and the effective
Lagrangian~\cite{heavylight}. As described in Sec.~\ref{sec:heft} we
use standard perturbation theory to compute a set of renormalized 1PI
Greens functions to determine the Wilson coefficients order by order
in the electroweak gauge coupling and in the mass dimension.

For the case of EW precision observables we rely on the renormalized
one-loop gauge boson vacuum polarizations in the UV complete model,
\begin{align}
\Pi^\text{(R)}_{VV}(p^2) = \Pi_{VV}(p^2) - (p^2- m_V^2)\, \delta Z_V + \delta m^2_V \; .
 \label{eq:fullPiVV}
\end{align}
The contributions to the un-renormalized $\Pi_{VV}$ may be a
combination of light, heavy, and mixed light-heavy field loops.  All
UV divergences are absorbed by the mass and wave function
counter terms $\delta m_V^2,\delta Z_V$. In this expression we take
the limit of large heavy masses and expand $\Pi^\text{(R)}_{VV}(p^2)$
in powers of $v/\Lambda$.

In complete analogy, we compute $\Pi_{VV}$ as a function of the Wilson
coefficients in the effective Lagrangian. Also here we can have two
types of contributions: light particle loops including effective
couplings induced at tree level, and tree-level insertions of loop
induced operators.  While the former are in general UV divergent, the
latter include the relevant counter terms of the Wilson
coefficients. Finally, we identify the two renormalized expressions
for the Greens function and determine the finite parts of the
renormalized Wilson coefficients. We will illustrate for our different
examples how this matching procedure based on Greens functions
leads to additional contributions to the Wilson coefficients compared to the functional approach.\bigskip

From the field theory point of view, it is interesting to compare this
approach to a leading-log resummation: the scale dependence $c_i(\mu)$
is the key element which relates the high-scale and low-scale regimes
in the effective Lagrangian. We always start with the effective
Lagrangian in the unbroken high-scale phase and for instance integrate
out the gauge invariant and separately renormalizable heavy
sector/heavy fields. Then we match the effective Lagrangian to the
full model at the intrinsic heavy mass scale $\Lambda =
\sqrt{2\lambda_2\,v_s^2}$. To get to the physically relevant energy
scales we can either evolve the running Wilson coefficients to the low
scale below electroweak symmetry breaking; or we introduce explicit
counterterms for the Wilson coefficients entering the low-scale Greens
functions, mixing the different dimension-6 operators.  Both ways
consistently extend the effective Lagrangian to the broken phase and
have to lead to the same predictions --- to leading-log accuracy in
the first case and exactly to fixed order in the second.

%%%%%%%%%%%%%%%%%%%%%%%%%%%%%%%%%%%%%%%%%%%%%%%%%%%%%%%%%%%%
\section{Oblique electroweak precision parameters}
\label{sec:oblique}

Effects of new physics on the electroweak gauge sectors can be
approximately described in terms of oblique parameters. The two most
relevant parameters constraining large classes of models are
\begin{align}
 \frac{\aem}{4 s_w^2 c_w^2} S &= \left[ - \Pi_{\gamma \gamma}' + \Pi_{ZZ}' - \Pi_{\gamma Z}' \dfrac{c_w^2 - s_w^2}{c_w s_w}  \right] - 
  \Big[ \cdots \Big]_\text{SM} 
\notag \\[4mm]
 \aem T &= \left[ \Pi_{WW} - \Pi_{ZZ} \right] - 
  [ \cdots ]_\text{SM} \; .
\label{eq:st_def}
\end{align}
The self energies $\Pi^{(\prime)}$ are evaluated at zero momentum
transfer and can be defined by the dimension-4 Lagrangian
\begin{align}
\lag = \lag_\text{SM} 
&- \frac{\Pi'_{\gamma \gamma}}{4} {F}_{\mu \nu} {F}^{\mu \nu} 
- \frac{\Pi'_{WW}}{2} {W}_{\mu \nu} {W}^{\mu \nu} 
- \frac{\Pi'_{ZZ}}{4} {Z}_{\mu \nu} {Z}^{\mu \nu} 
- \frac{\Pi'_{\gamma Z}}{4} {F}_{\mu \nu} {Z}^{\mu \nu} \notag \\
&- \Pi_{WW} \, {m}_W^2 {W}^+_\mu {W}^-{}^\mu 
- \frac{\Pi_{ZZ}}{2} \, {m}_Z^2 {Z}_\mu {Z}^\mu \; ,
\label{eq:st_lag}
\end{align}
using the standard definitions of the field strengths after
electroweak symmetry breaking. Some contributions to $\Pi^{(\prime)}$
are already induced through Standard Model loops, which are removed
through the above definition of the parameters $S$ and $T$. 

The self energy diagrams are by definition evaluated at zero momentum
transfer, which means that the only scales which enter are the weak
gauge boson and Higgs boson masses and, in case of new physics
contributions, the masses of the new particles. 

%%%%%%%%%%%%%%%%%%%%%%%%%%%%%%%%%%%%%%%%%%%%%%%%%%%%%%%%%%%%
\subsection{Effective Lagrangian}
\label{sec:oblique_eft}

%-----------------------------------------------------
\begin{figure}[t]
\includegraphics[width=0.7\textwidth]{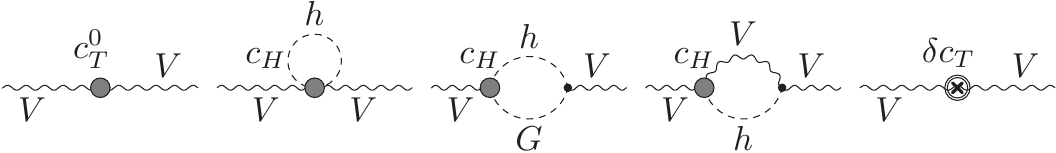}
\caption{Generic Feynman diagrams for the weak boson self-energies
  using a dimension-6 effective Lagrangian~\cite{feynarts}.  The shaded dots denote
  the dimension-6 operators.}
\label{fig:self-portal-eft}
\end{figure}
%-----------------------------------------------------

The definition of the oblique parameters Eq.\eqref{eq:st_lag} and some
terms in the effective Lagrangian of Eq.\eqref{eq:EFT} are similar in
structure. Comparing them we see that new physics effects in the
dimension-6 Lagrangian contribute to the $\Pi^{(\prime)}$ in two ways: 
First, loop-induced dimension-6 operators directly contribute
to $S$ and $T$ at tree level. Some corresponding Feynman diagrams are
shown in Fig.~\ref{fig:self-portal-eft}. The corresponding Wilson coefficients are
of the order $\aem$, and we have to include the appropriate
counter terms to renormalize the Wilson coefficients.  In the operator
basis of Eq.\eqref{eq:EFT} the Wilson coefficient $\wilson_T$ is
responsible for $T$, while a combination of $\wilson_B$ and
$\wilson_W$ generates a non-zero $S$ parameter,
\begin{align}
\aem T \Bigg|_\text{tree insertion} &= \wilson_T \frac{v^2}{\Lambda^2} \,,
& 
S \Bigg|_\text{tree insertion} &= 4 \pi \frac{v^2}{\Lambda^2} \,\left( \wilson_W + \wilson_B  \right) \, .
\label{eq:tpara_tree}
\end{align}
Second, we can insert operators generated at tree level into Standard
Model one-loop diagrams, as shown in Fig.~\ref{fig:self-portal-eft}.
These operators obviously only exist for models where there are
relevant electroweak Wilson coefficients induced at tree level. In
that case their contributions to $S$ and $T$ are also of order
$\aem/\Lambda^2$.  To illustrate their structure we look at a generic
tree-level-induced Wilson coefficient $\wilson_H$. The two relevant
self energy contributions for $T$ in the limit of zero momentum
transfer have similar structures,
\begin{align}
\Pi_{WW} 
&= \frac{\aem\,\wilson_H v^2}{16\pi\swd \Lambda^2}\, \left[
  3 \uvpole 
- 4 \log \frac{\mhd}{\mu^2}
+ \frac{5}{2} -\frac{\mhd}{2 \mwd} 
+ \log\frac{\mwd}{\mu^2} 
- \frac{4\mwd-\mhd}{\mhd-\mwd}\, \log\frac{\mhd}{\mwd}
 \right] \notag \\
\Pi_{ZZ} 
&= \frac{\aem\,\wilson_H v^2}{16\pi\swd \cwd\, \Lambda^2} \, \left[ 
  3 \uvpole 
- 4 \log\frac{\mhd}{\mu^2}
+ \frac{5}{2} - \frac{\mhd}{2 \mzd} 
+ \log\frac{\mzd}{\mu^2} 
- \frac{4\mzd-\mhd}{\mhd-\mzd}\, \log\frac{\mhd}{\mzd}
 \right] \; ,
\label{eq:pieft-zero}
\end{align}
with $\uvpole \approx 2/(4-n)$ describing the ultraviolet
divergence. This divergence needs to be absorbed by renormalizing the Wilson coefficients $\wilson_T$ in the tree-level relation in Eq.\eqref{eq:tpara_tree}. The counter term of $c_T$ or, alternatively, its anomalous
dimension can be linked to $\wilson_T$--$\wilson_H$ operator mixing at
one loop and is of the form $\delta \wilson_T \propto \aem
\wilson_H$. Altogether, the ultraviolet poles cancel and all contributions
combine to a finite contribution
\begin{align}
\aem T \Bigg|_\text{weak loops}
 = \frac{3\aem \, \wilson_H v^2}{16\pi\swd\,m_W^2\, \Lambda^2}\, \left[
\frac{m_Z^2\,\mhd}{\mhd-m_Z^2}\,\log \frac{\mhd}{m_Z^2} - \frac{m_W^2\,\mhd}{\mhd-m_W^2}\,\log \frac{\mhd}{m_W^2} \right]\, ,
\label{eq:tpara_portal_eft}
\end{align}
which breaks custodial symmetry.  The explicit logarithms induced by
the weak-scale loops are of the kind $\log (m_h/m_V)$ with $V = W,Z$,
and $m_h = 126$~GeV denoting the Standard Model Higgs mass. They are
indicative of the aforementioned violation of custodial symmetry:
while at tree level $\wilson_H$ does not violate custodial symmetry,
at loop level it can mix with the other Wilson coefficients like
$\wilson_T$ through its anomalous dimension. The resulting
contribution to the $T$ parameter will then be proportional to the
logarithm of the scale splitting.\bigskip

As discussed in Sec.~\ref{sec:brokenphase}, at this order in 
perturbation theory we need to include both,
 weak-scale loops combined with tree-level operators and
tree-level diagrams with loop-induced operators.  They arise at the
same level of perturbation theory from the same dimension-6
Lagrangian. In contrast, a finite contribution to $U$ only occurs once
we include operators of higher mass dimension. The renormalization
scale of Wilson coefficients is naturally chosen to be of the order of the
electroweak scale, $\mu \sim m_W \sim m_Z \sim m_h \sim v$.  In our numerical
evaluation we fix $\mu = m_W$. For high-scale matching, the
renormalization scale dependence defines a leading-log approximation
of the Wilson coefficients in terms of $\aem \log (\Lambda/m_W)$~\cite{llew}.
This implies that dimension-6
contributions to both oblique parameters have leading contributions of the kind
\begin{align}
\aem T 
\sim \wilson_j(\mu = m_W) \, \frac{\aem v^2}{\Lambda^2} \, \log \frac{\Lambda^2}{m_W^2}
\sim \aem S \; ,
\end{align}
as well as sub-leading contributions without this, often not very
large, logarithm.

%%%%%%%%%%%%%%%%%%%%%%%%%%%%%%%%%%%%%%%%%%%%%%%%%%%%%%%%%%%%
\subsection{Higgs singlet extension}
\label{sec:oblique_portal}

A Higgs singlet extension or (renormalizable) Higgs portal is defined
by the extended scalar potential
\begin{align}
V(\phi,S) = 
  \mu^2_1\,(\phi^\dagger\,\phi) 
+ \lambda_1\,|\phi^{\dagger}\phi|^2 
+ \mu^2_2\,S^2
+ \lambda_2\,S^4 
+ \lambda_3\,|\phi^{\dagger}\,\phi|S^2 \; .
\label{eq:portal-potential}
\end{align}
It contributes to the oblique parameters in both ways described in
Sec.~\ref{sec:oblique_eft} because of the linear coupling of the heavy
scalar to the light Higgs doublet, $\lag \supset \lambda_3 v_s S \,
\phi^\dagger\phi$, with $\langle S \rangle =
v_s/\sqrt{2}$~\cite{singlet_ewpo}. The interplay between the
tree-level insertion of loop-induced operators and the loop-insertion
of tree-level-induced operators is the reason why this model is interesting to study. The
Higgs portal coupling $\lambda_3$ induces a finite mixing angle
angle~\cite{Brehmer:2015rna}
\begin{align}
\frac{\tan^2 (2\alpha)}{4}
= \frac{1}{4} \; \left( \frac{\lambda_3 v v_s}{\lambda_2 v_s^2 - \lambda_1 v^2} \right)^2 
\stackrel{v \ll v_s}{\approx} 
\frac{\lambda_3^2}{4 \lambda_2^2}\frac{v^2}{v_s^2} 
\approx \frac{\lambda_3^2}{2 \lambda_2}\,\frac{v^2}{\mHHd}
\approx \sad 
\qquad \text{with} \quad 
 \mHHd \approx 2\lambda_2\,v_s^2 \; ,
\label{eq:portal-angle}
\end{align}
from the interaction eigenstates to the two mass eigenstates $h$ and
$H$. All approximations indicated by `$\approx$' are leading in terms of
$v^2/\Lambda^2$. Explicit contributions from the new scalar are
proportional to $s_\alpha^2 \equiv \sin^2 \alpha$, while the modified
contribution from the Standard Model Higgs come with $c_\alpha^2-1 =  -
s_\alpha^2$.  This
means that the oblique parameters have the particularly simple leading
behavior~\cite{hlm,heft_limitations2}
\begin{align}
S &\approx \frac{\sad}{12\pi}\, \left( - \log\, \frac{\mhd}{\mzd} +  \log\, \frac{\mHHd}{\mzd} \right) 
   \approx \frac{\lambda_3^2}{24\pi \lambda_2}
         \,\frac{v^2}{\mHHd} \log\, \frac{\mHHd}{\mhd} \notag \\
%   \left[ 1 + \mathcal{O}\left( \frac{v^2}{\mHHd}\right) \right] 
T &=  \frac{-3\sad}{16\pi\, \swd \, \mwd}
      \left( \mzd \log\, \frac{\mHHd}{\mhd} 
           - \mwd \log\, \frac{\mHHd}{\mhd} \right) 
   \approx \frac{-3 \lambda_3^2 v^2}{32\pi\, \swd \, \lambda_2\, \mwd } \,
     \left( \frac{\mzd}{\mHHd}
           - \frac{\mwd}{\mHHd}  \right) \, \log\, \frac{\mHHd}{\mhd} \; .
%   \left[ 1 + \mathcal{O}\left(\frac{v^2}{\mHHd}\right) \right] 
\label{eq:st-portal-heavy}
\end{align}
While decoupling is guaranteed by the suppressed mixing angle
$s_\alpha \propto 1/m_H$, we notice the additional logarithms $\log
(m_H/m_h)$, which delay the decoupling of the heavy scalar.  In the
absence of a large hierarchy of scales we will see that the matching
of the full theory to the truncated Lagrangian is not uniquely
defined.  For each of the benchmark models in
Tab.~\ref{tab:portal-benchmarks} we evaluate the $S$ and $T$
parameters from the full model and from the effective Lagrangian,
considering different setups:

\begin{itemize}
\item LL-L: \textit{leading-log, loop-induced Wilson coefficients,}
  where we limit ourselves to the tree-level insertion of loop-induced
  operators $\oT$ and $\obw$ in the leading-log
  approximation~\cite{Chiang:2015ura}. Because for testable models the
  logarithm $\log (\Lambda/m_W)$ cannot be too large, we expect this
  approximation to not work too well.  If we choose the matching scale
  as $\Lambda = \sqrt{2\lambda_2 v_s^2}$ we find the Wilson
  coefficients
\begin{align}
 \frac{\wilson_T(\mu)}{\Lambda^2} =  - \frac{3\aew s_w^2 \lambda_3^2}{32 \pi c_w^2 \lambda_2 \Lambda^2} \,\log \frac{\Lambda^2}{\mu^2}
 \qqquad 
 \frac{\wilson_{B,W}(\mu)}{\Lambda^2} =  \frac{\lambda_3^2}{192 \pi^2 \lambda_2 \Lambda^2} \,\log \frac{\Lambda^2}{\mu^2} \; .
\label{eq:singlet-default}
\end{align}
  As discussed before, the logarithmic structure of $\wilson_{T,B,W}$
  follows from the specific way of breaking custodial symmetry.

\item LL-TL: \textit{leading-log, loop-induced Wilson coefficients
  plus weak-scale loops,} where we add the tree-level induced operator
  $\oh$ to the weak-scale loops. These correspond to the SM-like Higgs
  mediated contributions to the gauge boson polarization, with
  rescaled Higgs--gauge boson couplings. In addition to the Wilson
  coefficients given in Eq.\eqref{eq:singlet-default} we find
\begin{align}
 \frac{\wilson_H}{\Lambda^2} = \frac{\lambda_3^2}{2\lambda_2 \Lambda^2} \; .
\label{eq:singlet-default2}
\end{align}

\item LL-TL$v$: \textit{$v$-improved leading-log, loop-induced Wilson
  coefficients plus weak-scale loops,} where we adjust the matching
  procedure to include $v$-induced terms~\cite{Brehmer:2015rna}. The
  matching scale is shifted to the mass of the new state,
  $\Lambda = \mHH$. In addition, we express the full model predictions
  in terms of the mixing angle $c_\alpha$, so the corresponding Wilson coefficients
  become 
\begin{align}
 \frac{\wilson_H}{\Lambda^2} = \frac{2(1-c_\alpha)}{v^2}
 \qquad
 \frac{\wilson_T(\mu)}{\Lambda^2} =  - \frac{3\aew s_w^2 \,(1-c_\alpha)}{8 \pi c_w^2\,v^2}\,\log \frac{m_H^2}{\mu^2}
 \qqquad 
 \frac{\wilson_{B,W}(\mu)}{\Lambda^2} =  \frac{1-c_\alpha}{48 \pi^2\,v^2}\,\log \frac{m_H^2}{\mu^2} \; .
\label{eq:singlet-vimproved}
\end{align}
  The explicit scale suppression is now replaced by the dependence on
  the mixing angle with $1- c_\alpha \approx \lambda_3^2 v^2/(4
  \lambda_2^2 v_s^2)$, neglecting higher powers of $v/v_s$.  This
  modification with respect to the default matching is equivalent to
  resumming part of the higher-dimensional Higgs vev insertions.

\item BP-TL: \textit{broken-phase matching, loop-induced Wilson
  coefficients plus weak-scale loops,} where unlike in the LL-TL
  scheme we now perform the matching with the full operators in the
  broken phase. The matching based on Greens functions does not change the
  Wilson coefficient $\wilson_H$ entering at tree level, but the
  loop-induced operators which until now are only included with their
  leading logs. Choosing the default matching scale $\Lambda =
  \sqrt{2\lambda_2 v_s^2}$ we find
\begin{align}
 \frac{\wilson_T(\mu)}{\Lambda^2} = -\frac{\aew s_w^2 \lambda_3^2}{32 \pi c_w^2 \lambda_2 \Lambda^2}
                    \left( -\frac{5}{2}+ 3\log \frac{\Lambda^2}{\mu^2} \right)
 \qqquad 
 \frac{\wilson_{B,W}(\mu)}{\Lambda^2} = \frac{\lambda_3^2}{576 \pi^2 \lambda_2 \Lambda^2}
                      \left( -\frac{5}{2} + 3\log \frac{\Lambda^2}{\mu^2}\right)  \; .
 \label{eq:singlet-explicit}
\end{align}
  Compared to the LL-L result in Eq.~\eqref{eq:singlet-default}, we
  obtain a finite term $-5/2$ at order $\mathcal{O}(v^2/\Lambda^2)$ as
  the genuine contribution from the explicit broken-phase matching.

\item BP-TL$v$: \textit{$v$-improved broken-phase matching,
  loop-induced Wilson coefficients plus weak-scale loops,} where we
  apply the $v$-improved matching prescription to the BP-TL setup.
  Both the explicit matching and the $v$-improvement can now be
  regarded as strategies to incorporate vev-dependent corrections to
  the default ideal(ized) effective theory. Combining all improvements
  we find for the Wilson coefficients
\begin{align}
 \frac{\wilson_T(\mu)}{\Lambda^2} = -\frac{\aew s_w^2 (1-c_\alpha)}{8 \pi c_w^2 v^2}
               \left( -\frac{5}{2}+ 3\log\frac{m_H^2}{\mu^2} \right)
 \qqquad 
 \frac{\wilson_{B,W}(\mu)}{\Lambda^2} = \frac{1-c_\alpha}{144 \pi^2 v^2}\,
                \left( -\frac{5}{2}+3\log\frac{m_H^2}{\mu^2}\right) \; .
 \label{eq:singlet-explicitimproved}
\end{align}
  This result systematically includes the numerically relevant
  corrections of order $v/\Lambda$ to the usual matching scheme.

\end{itemize}
\bigskip

%-----------------------------------------------------
\begin{table}[t]
\begin{tabular}{l|lllllll|ll}
\hline  
& $\mHH$ & $\sa$ & $\tan\beta$ & $\Lambda = \sqrt{2\lambda v_s^2}$ & $\lambda_1$ & $\lambda_2$ &$\lambda_3$ & \multicolumn{2}{c}{$c_H v^2/\Lambda^2$} \\ 
& & & & &  & & & LL-TL &  BP-TL$v$  \\
\hline
S1 & 300 & 0.1&  10 &   298.8 & 0.13 & 7.1$\times 10^{-3}$ & 1.2$\times 10^{-2}$ & 6.9$\times 10^{-3}$ & 1.0$\times 10^{-2}$ \\ 
S2 & 700 & 0.1&  10 &   696.6 & 0.16 & 3.9$\times 10^{-2}$ & 7.5$\times 10^{-2}$ & 9.5$\times 10^{-3}$ & 1.0$\times 10^{-2}$ \\ 
S3 & 300 & 0.3&  10 &   288.6 & 0.18 & 6.6$\times 10^{-3}$ & 3.4$\times 10^{-2}$ & 6.5$\times 10^{-2}$ & 9.2$\times 10^{-2}$ \\ 
S4 & 500 & 0.3&  10 &   668.8 & 0.46 & 3.6$\times 10^{-2}$ & 2.2$\times 10^{-1}$ & 9.2$\times 10^{-2}$ & 9.2$\times 10^{-2}$ \\ \hline
\end{tabular}
\caption{Benchmark points for the Higgs portal. The Wilson coefficient
  $\wilson_H$ is given in the default and the $v$-improved schemes.
  All mass scales are given in GeV.}
\label{tab:portal-benchmarks}
\end{table}
%-----------------------------------------------------

In Tab.\ref{tab:portal-benchmarks} we introduce a set of benchmark
points for the Higgs portal scenario, defining two trajectories in
$m_H$ for given $\tan \beta = 10$ and $s_\alpha = 0.1$ or $s_\alpha =
0.3$.  The self-couplings are related to the mixing angles through
\begin{align}
\sad = \frac{\mhd-2\lambda_1\,v^2}{\mhd-\mHHd}\; ; \qquad \qquad 
\tan^2\beta = \frac{v_s^2}{v^2} = \frac{\mhd+\mHHd-2\lambda_1 v^2}{2\lambda_2\,v^2} \; .
\label{eq:angles-decoupling}
\end{align}
In the well-known decoupling limit~\cite{decouple} the two angles are related as
$s_\alpha^2 \sim 1/(1 + \tan^2 \beta)$ and for example give $s_\alpha \to
0$ together with $\beta \to \pi/2$.

%-----------------------------------------------------
\begin{table}[b!]
 \begin{tabular}{l|l|rrrrrr} \hline
&  & full model & LL-L & LL-TL & LL-TL$v$ & BP-TL & BP-TL$v$ \\ \hline   
\multirow{4}{*}{$S$} & S1 &  $6.22\times10^{-4}$ &  $4.79\times10^{-4}$ &  $6.26\times10^{-4}$ &  $9.15\times10^{-4}$ &  $4.74\times10^{-4}$ &  $6.94\times10^{-4}$  \\ 
                     & S2 &  $1.13\times10^{-3}$ &  $1.08\times10^{-3}$ &  $1.29\times10^{-3}$ &  $1.37\times10^{-3}$ &  $1.08\times10^{-3}$ &  $1.14\times10^{-3}$ \\   
                     & S3 &  $5.60\times10^{-3}$ &  $4.43\times10^{-3}$ &  $5.83\times10^{-3}$ &  $8.41\times10^{-3}$ &  $4.38\times10^{-3}$ &  $6.38\times10^{-3}$ \\ 
                     & S4 &  $1.01\times10^{-2}$ &  $1.04\times10^{-2}$ &  $1.23\times10^{-2}$ &  $1.26\times10^{-2}$ &  $1.03\times10^{-2}$ &  $1.05\times10^{-2}$\\ \hline
\multirow{4}{*}{$T$} & S1 & $-8.30\times10^{-4}$ & $-1.39\times10^{-3}$ & $-8.07\times10^{-4}$ & $-1.18\times10^{-3}$ & $-3.67\times10^{-4}$ & $-5.41\times10^{-4}$\\
                     & S2 & $-1.93\times10^{-3}$ & $-3.14\times10^{-3}$ & $-2.34\times10^{-3}$ & $-2.49\times10^{-3}$ & $-1.74\times10^{-3}$ & $-1.85\times10^{-3}$  \\
                     & S3 & $-7.47\times10^{-3}$ & $-1.28\times10^{-2}$ & $-7.32\times10^{-3}$ & $-1.09\times10^{-2}$ & $-3.14\times10^{-3}$ & $-4.97\times10^{-3}$\\
                     & S4 & $-1.74\times10^{-2}$ & $-3.00\times10^{-2}$ & $-2.22\times10^{-2}$ & $-2.29\times10^{-2}$ & $-1.63\times10^{-2}$ & $-1.70\times10^{-2}$\\ \hline
  \end{tabular}
 \caption{Predictions for $S$ and $T$ in the singlet extension for the
   full model and the different effective Lagrangian setups.  The
   benchmark points are defined in
   Tab.~\ref{tab:portal-benchmarks}.}
 \label{tab:singlet-stu}
\end{table}
% -------------------------------------------------------------------

In Table ~\ref{tab:singlet-stu} we evaluate the $S$ and $T$ parameters
for the full model and confront the results with the different
matching schemes. Default, leading-log matching in the unbroken phase
(LL-L) essentially does not reproduce the full model, and even
$v$-improvement (LL-TL$v$) still leads to a poor agreement with the
full prediction.  The most accurate BP-TL$v$ setup agrees with the
full model typically within a few per-cent.  However, we also find
sizable discrepancies of up to $\mathcal{O}(30)\% $ for the points S1
and S3, where the relatively low heavy singlet mass ruins the scale
hierarchy. \bigskip

%-----------------------------------------------------
\begin{figure}[t]
\includegraphics[width=0.4\textwidth]{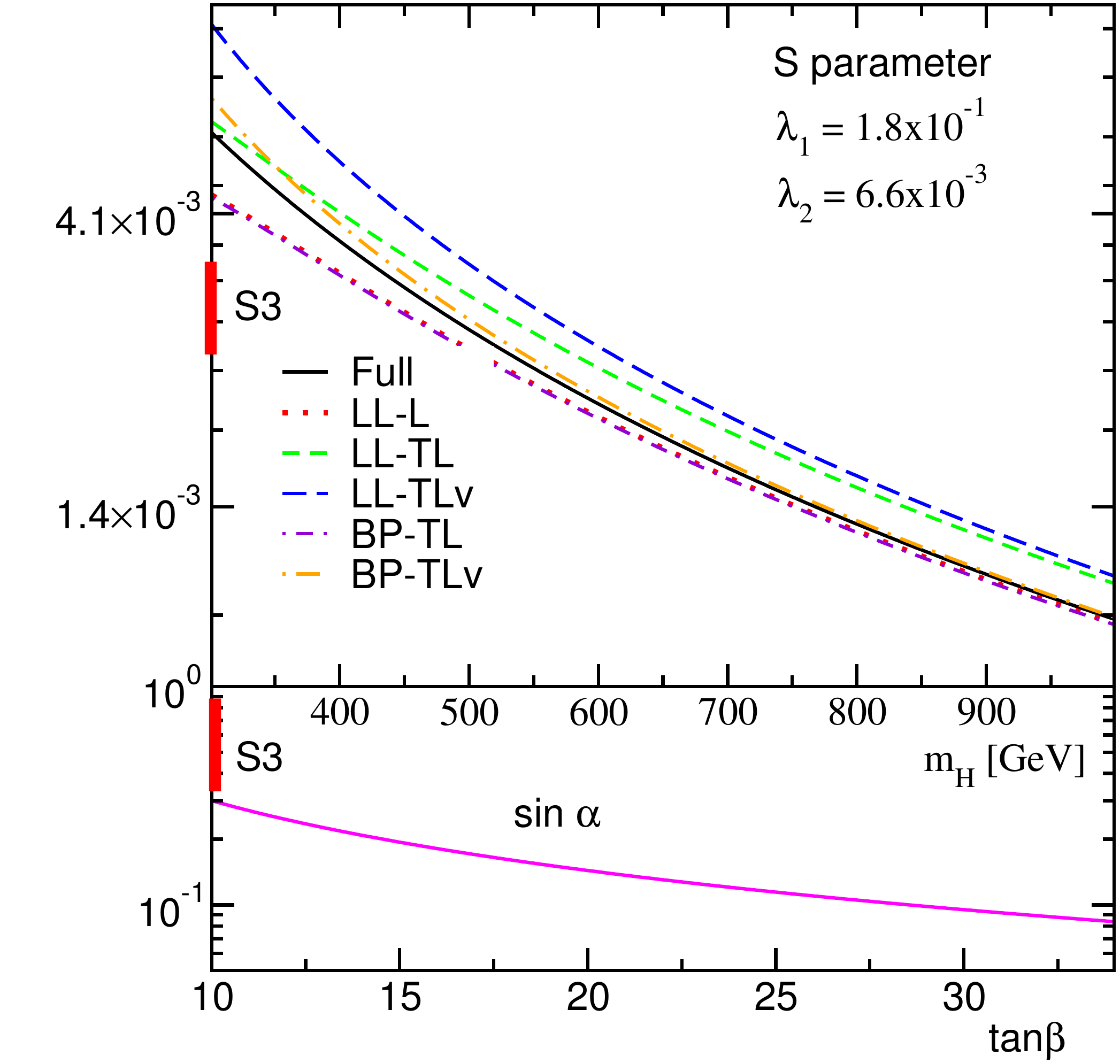}
\hspace*{0.08\textwidth}
\includegraphics[width=0.4\textwidth]{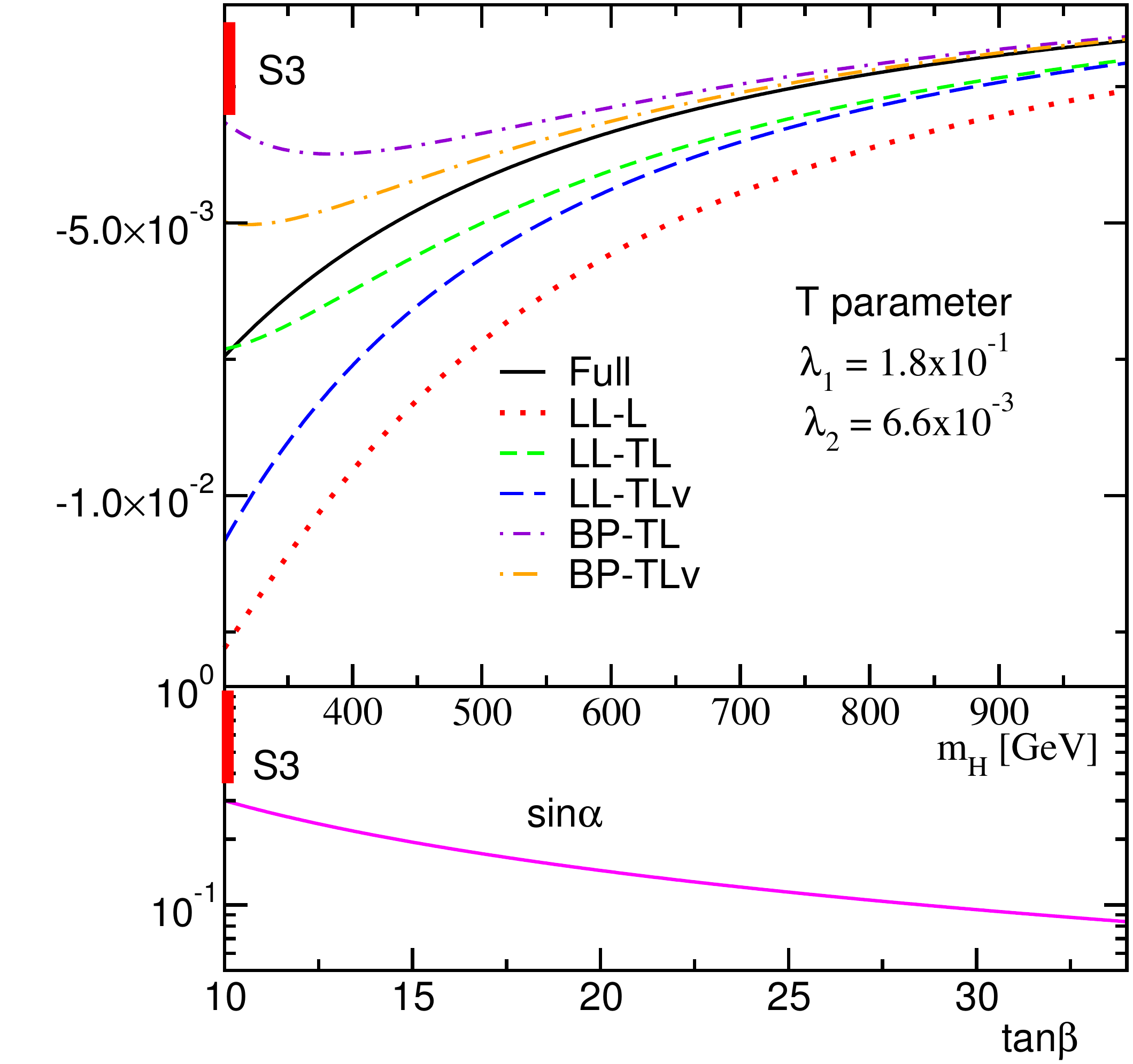} 
\caption{Dependence of $S$ (left) and $T$ (right) on the heavy scalar
  mass $m_H \approx \Lambda$ towards decoupling. We fix the
  self-couplings $\lambda_{1,2}$ and compute the mixing angles
  $\alpha$ and $\beta$ following Eq.\eqref{eq:angles-decoupling} from
  the varying heavy Higgs mass.  The small panels give $s_\alpha$ as a
  function with $\tan\beta = v_s/v$. The red bar indicates the S4
  benchmark point from Tab.~\ref{tab:portal-benchmarks}.}
 \label{fig:ST-singlet-decoupling}
\end{figure}
%-----------------------------------------------------

In Fig.~\ref{fig:ST-singlet-decoupling} we show the decoupling of the
oblique parameters for the Higgs portal model and its different
dimension-6 approximations.  From Eq.\eqref{eq:st-portal-heavy} we
know that both parameters will approach zero in the decoupling limit
with a quadratic power suppression, softened by a logarithm $\log
(m_H/m_h)$. The leading power suppression arises through the mixing
angle $\alpha$. To reflect this dependence we keep $\lambda_{1,2}$
constant and vary the heavy Higgs mass $m_H$.  Our starting
configuration is the benchmark point S4 with $s_\alpha = 0.3$,
$\tan\beta = 10$, and $m_H = 300$~GeV.  We then decouple the heavy
scalar by increasing the physical mass eigenvalue $m_H$. The lower
sub-panels in Fig.~\ref{fig:ST-singlet-decoupling} correlate the
variation of the two mixing angles, which obey $s_\alpha^2 = 1/(a +
b\tan^2 \beta)$, with $a,b$ being functions of $m_h$ and
$\lambda_{1,2}$.  In this situation, the very heavy extra scalar is
almost entirely singlet-like, and acquires its large mass through the
intrinsic singlet scale $|\mu^2_s| \sim \lambda_2\,v_s^2$.

For the $S$ parameter we first observe that not all of the effective
Lagrangian approximations give the correct decoupling pattern.
Skipping the naive LL-L for now, we see that the leading-log
approximation together with the weak-scale loops LL-TL as well as its
$v$-improved counter part LL-TL$v$ %fail. Both of them 
significantly
disagree with the full model towards large values of $m_H$.
The
situation only improves once we employ a broken-phase matching in our
BP-TL scheme, further enhanced significantly in its $v$-improved
version BP-TL$v$. This tells us that contributions beyond the plain
dimension-6 truncation lead to more accurate results only if the
complete $\ord(v)$ dependence is included. The fact that the simple LL-L
approach agrees very well with the full model has to be considered
accidental, and we will look at it again below.

Next, we notice that the effective Lagrangians describe the $T$
parameter significantly worse, in particular for small values of the
heavy Higgs mass $m_H$. The reason is its enhanced sensitivity to the
relative splittings between $m_W$, $m_Z$, and $m_h$, whereas $S$ really
only depends on $m_Z$. Among the different approximations we still see
that only BP-TL$v$ really describes the decoupling accurately,
in complete analogy to the $S$ parameter. Unlike for the $S$
parameter, the naive leading-log approximation LL-L exhibits
a poor performance, as expected.\bigskip

%-----------------------------------------------------
\begin{figure}[t]
\includegraphics[width=0.4\textwidth]{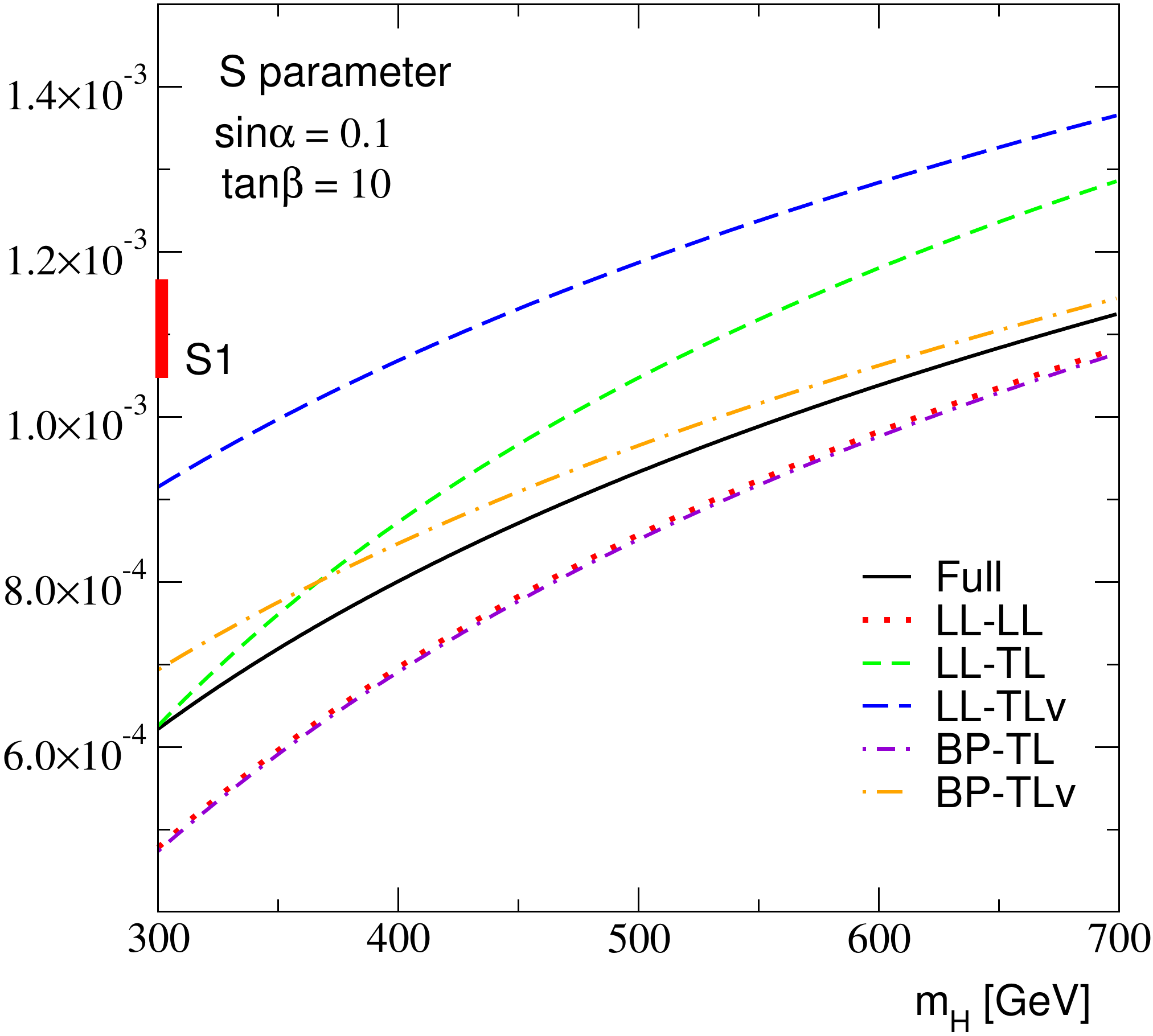}
\hspace*{0.08\textwidth}
\includegraphics[width=0.4\textwidth]{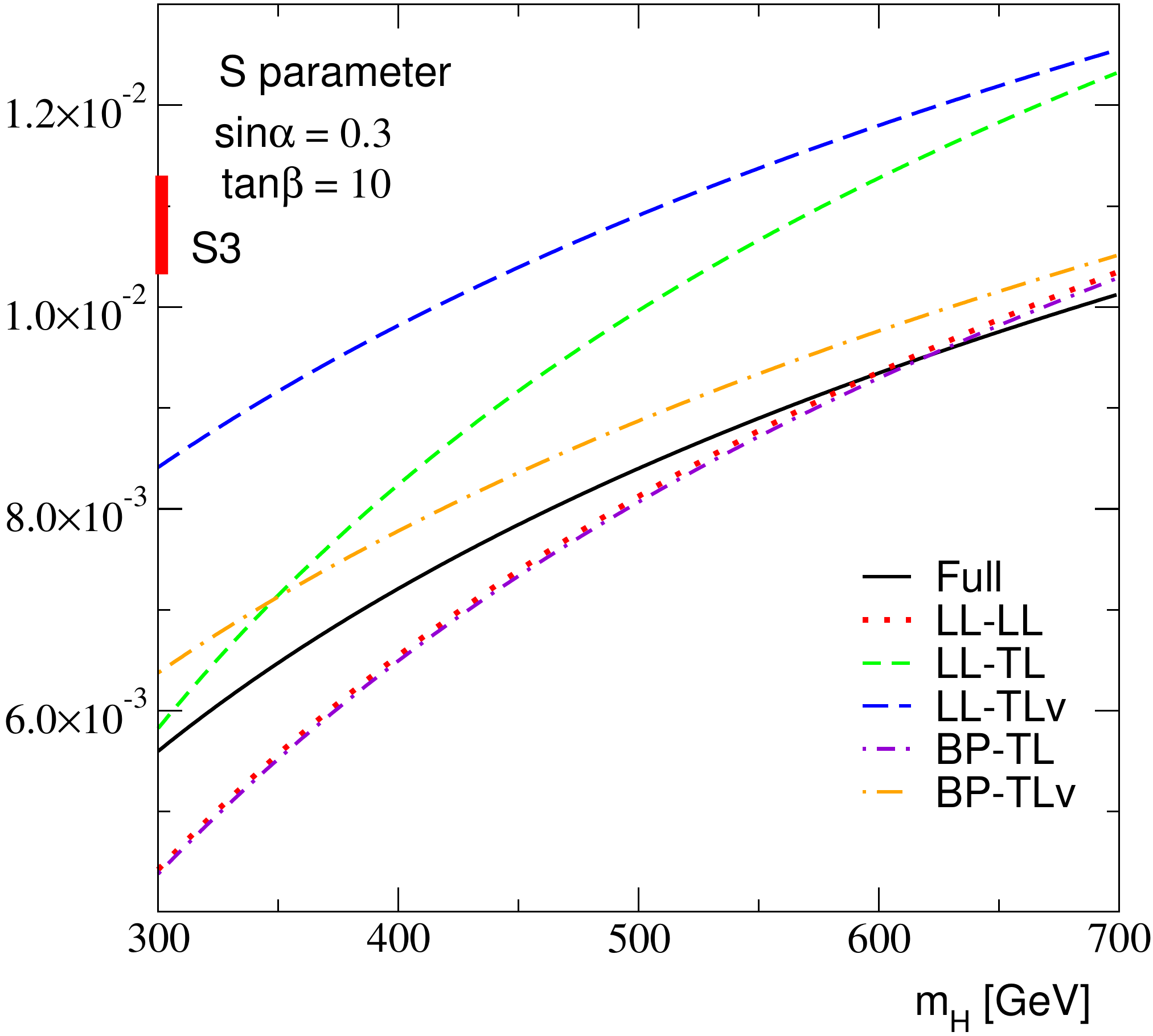}\\ 
\includegraphics[width=0.4\textwidth]{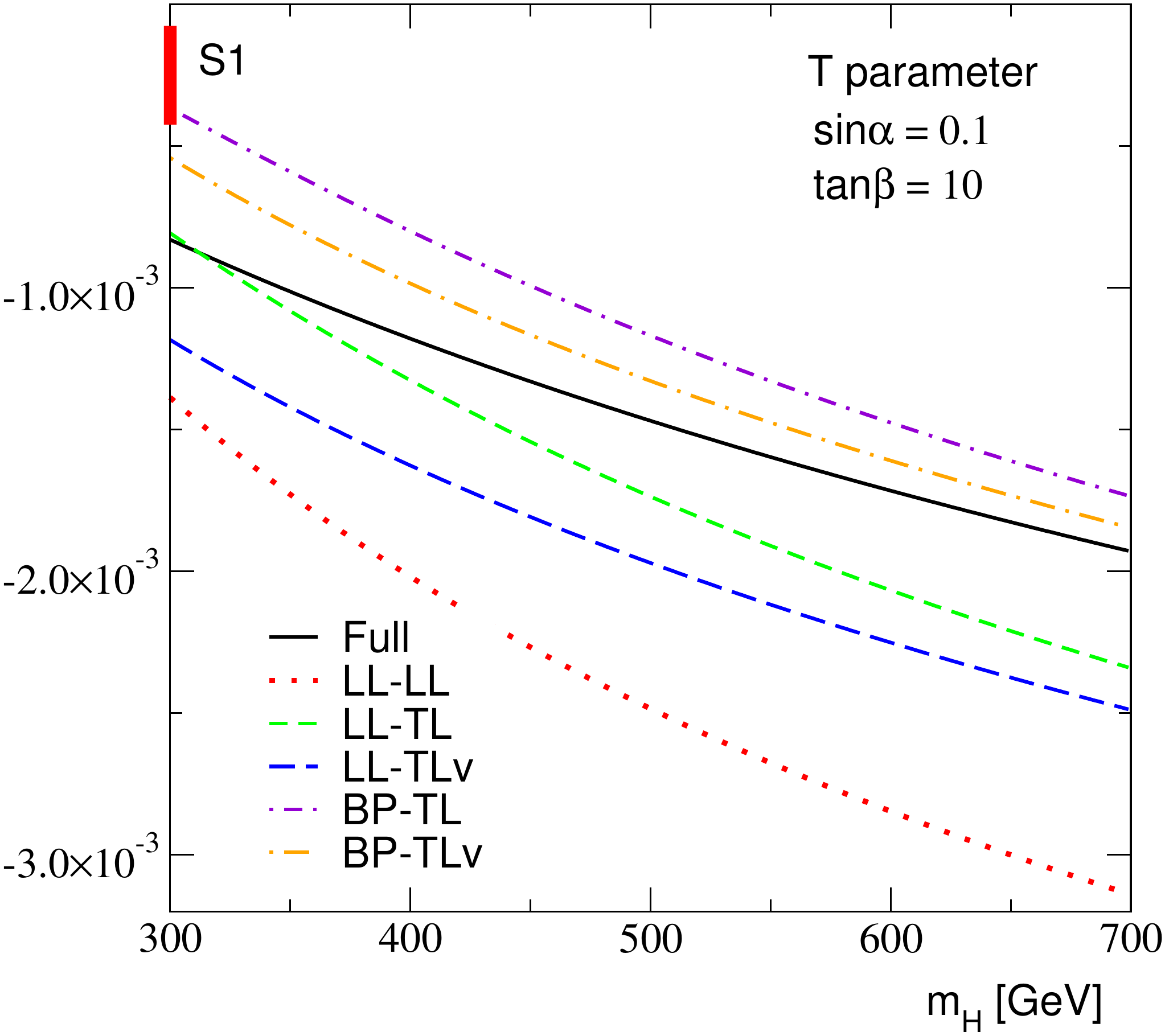}
\hspace*{0.08\textwidth}
\includegraphics[width=0.4\textwidth]{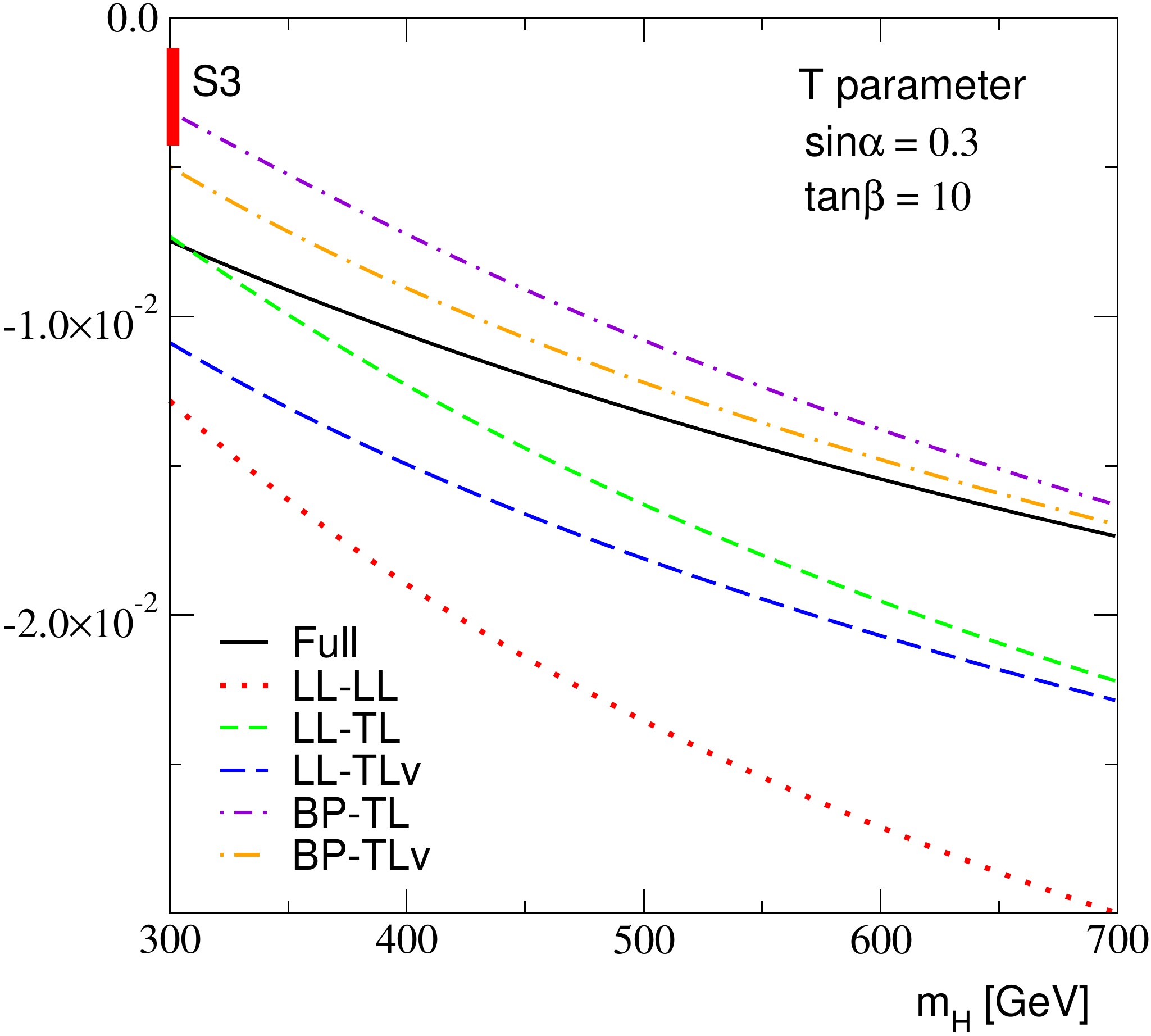}
\caption{Dependence of the $S$ (upper panels) and $T$ (lower panels)
  parameters on the heavy scalar mass $m_H \approx \Lambda$ for two
  choices of mixing angles. Because the mixing angle is kept constant
  in each panel we only see the logarithmic modulation of the
  decoupling, shown for example in Eq.\eqref{eq:st-portal-heavy}. The
  red bar indicates the benchmark points S1 (S4) from
  Tab.~\ref{tab:portal-benchmarks}.}
\label{fig:portal-overmass}
\end{figure}
%-----------------------------------------------------

The distinguishing feature of the oblique parameters in this model is their delayed
decoupling because of the additional logarithm. To study this
logarithmic behavior we show the three different definitions of the
dimension-6 predictions compared to the full model for fixed
$s_\alpha$ in Fig.~\ref{fig:portal-overmass}. In the upper panels we
show the $S$ parameter for the two different mixing angles.  Following
Eq.\eqref{eq:portal-angle} a larger mixing angle corresponds to a
weaker scale hierarchy $v/v_s$.  The benchmark points of
Tab.~\ref{tab:portal-benchmarks} correspond to the minimum and maximum
$m_H$ values.

Again skipping the LL-L setup for the $S$ parameter for now, we start
with the standard leading-log LL-TL scenario. While for small $m_H$
this matching scheme seems to agree very well with the full model, at larger $m_H$ we see that
this agreement is accidental: increasing the hierarchy of scales and
reducing the size of the perturbative parameter (in and beyond the
logarithm) makes things significantly worse. This is a clear
indication that the leading-log approximation of the Wilson
coefficients fails.  $v$-Improving the matching procedure for these
leading-log terms in the LL-TL$v$ scheme actually worsens the
agreement between the full model and the dimension-6
approximation. For both mixing angles it overshoots the full model
description by around 50\% for low $m_H$ and 20\% for high $m_H$. As
mentioned above, this suggests that in spite of a coincidentally good
agreement of the LL-L scheme with the full model the leading-log
approximation to the $S$ parameter fails systematically. The behavior
of the LL-L curve for $s_\alpha = 0.3$ confirms this picture, because
it crosses the full model curve rather than approximating it towards
larger $m_H$.

Extending the full LL-TL scheme to include some $v$-dependent terms
through broken phase matching has a sizeable effect on the effective
Lagrangian prediction, as was noted also in Ref.~\cite{nloewmatch}.
However, it again does not lead to a significant
improvement of the $m_H$ dependence, either. Only the
full set of operators computed without the leading-log approximation
and including $v$-induced effects through $v$-induced matching in the
broken phase leads to an agreement of the full model with the
effective Lagrangian at 10~...~15\% for low $m_H$ and at 2~...~4\% for
large $m_H$. A crucial consistency check is the appropriate
log-modulated decoupling behavior towards large $m_H$ values, which we
only observe for the BP-TL$v$ approach.

Moving on to the $T$ parameter we observe a similar behavior. First,
unlike in Fig.~\ref{fig:ST-singlet-decoupling} we do not observe any
curves turning over. Instead, we see that this feature is driven by a
very poor description of the logarithmic structure in many of the
effective Lagrangian approaches. As expected, the LL-L approach now
fails badly, adding the weak loops does not improve the too steep
dependence on $m_H$, and $v$-improvement alone does not help
either. Instead, only the properly matched and $v$-improved BP-TL$v$
scheme leads to an acceptable description of the delayed decoupling of
the heavy scalar.

As for the two different ways to improve the matching, we see that
$v$-improvement by itself does not improve the agreement between the
dimension-6 approximation and the full model.  This illustrates the
presence of different orders in the perturbative and the effective
Lagrangian expansions: $v$-improvement resums a subset of $d>6$
contributions in $(v/\Lambda)^{d-2}$.  When computing quantum
corrections, this approach is only meaningful if we first ensure that
no equally relevant higher-order terms in the perturbative expansion
are neglected. Applying $v$-improvement combined with full
broken-phase matching indeed reconciles the dimension-6 results with
the full model.

%%%%%%%%%%%%%%%%%%%%%%%%%%%%%%%%%%%%%%%%%%%%%%%%%%%%%%%%%%%%
\subsection{Scalar top partners}
\label{sec:oblique_partners}

New colored scalar particles are, strictly speaking, not an extension
of the SM Higgs sector, but they can lead to interesting modifications
of the LHC observables. We consider a scalar top-partner sector
mimicking the stop and sbottom sector of the MSSM. Its Lagrangian has
the form\footnote{Unlike for example in Ref.\cite{Brehmer:2015rna} we
  now define $\kLR$ with a mass dimension, because its potential
  suppression scale is not uniquely defined once we include different
  loop-level matching schemes.}
\begin{align}
 \lag \supset& \,  (D_{\mu}\,\Qtilde)^\dagger\,(D^\mu\Qtilde) + (D_\mu\,\TR)^*\,(D^\mu\,\TR)
 - \tilde{Q}^\dagger\, \MLeft^2\,\tilde{Q}\, - \MRight^2\,\TR^*\,\TR \notag \\
&-\kLL\,(\phi \cdot \Qtilde)^\dagger\,(\phi \cdot\Qtilde) -\kRR\,(\TR^*\TR)\,(\phi^\dagger\,\phi) 
 - \left[ \kLR \, \TR^*\,(\phi \cdot \Qtilde) + \text{h.c.} \right]
 \label{eq:partner_lag}.
 \end{align}
Here, $\Qtilde$ and $\TR$ are the additional isospin doublet and
singlet in the fundamental representation of $SU(3)_C$.  The singlet
state $\sbR$ is assumed to be heavier and integrated out.  This leaves
us with three physical degrees of freedom, the scalars $\tilde{t}_1$,
$\tilde{t}_2$ and $\tilde{b}= \tilde{b}_L$.  The reminiscent
underlying R-parity precludes any linear coupling involving the heavy
fields. Thus there are no contributions to the oblique parameters proportional
to a tree-induced Wilson coefficient inside weak-scale loops. The
non-Higgs scalar mass matrix has the form
\begin{align}
\begin{pmatrix}
\MLeft^2+ \kLL \dfrac{v^2}{2} & \kLR\,\dfrac{v}{\sqrt{2}} \\
\kLR\,\dfrac{v}{\sqrt{2}} & \MRight^2+ \kRR \dfrac{v^2}{2} 
\end{pmatrix} \; .
\label{eq:partner_matrix}
\end{align}
After diagonalization we can write the physical top partner masses
such that they reflect the scale hierarchies of the effective Lagrangian,
\begin{align} 
\mst{1}^2 
&= \MLeft^2\ctd + \MRight^2\,\std  + \frac{v^2}{2} \left( \kLL\ctd +\kRR\std + \frac{\sqrt{2}\kLR}{v}\sdt\right)
\approx  M^2 + \frac{v^2}{2} \left( \kLL\ctd + \kRR\std + \frac{\sqrt{2}\kLR}{v} \sdt \right) \notag \\
\mst{2}^2 
&= \MLeft^2\std + \MRight^2\,\ctd  + \frac{v^2}{2} \left( \kLL\std +\kRR\ctd - \frac{\sqrt{2}\kLR}{v}\sdt\right)
\approx M^2 + \frac{v^2}{2} \left( \kLL\std + \kRR\ctd - \frac{\sqrt{2}\kLR}{v} \sdt \right)  \notag \\
\msb{}^2 
&= \MLeft^2 \approx M^2 \; ,
\label{eq:spartner-masses-redefined}
\end{align}
where $\st = \sin \theta_{\tilde{t}}$ and $\ct = \cos
\theta_{\tilde{t}}$, etc.  The stop mixing angle itself also depends
on $M$ and the $\kappa_j$, but we keep it in
Eq.\eqref{eq:spartner-masses-redefined} in the interest of a compact
formula. Also shown are the simplified results when assuming a single
heavy mass scale $M \equiv \MLeft = \MRight$. Independently of this
approximation, the physical mass eigenstates exhibit a mass splitting
of $\ord(v^2/M^2)$ after electroweak symmetry breaking. A detailed
description of the model can be found in the Appendix of
Ref.~\cite{Brehmer:2015rna}.\bigskip

Assuming, in addition, small mixing $\st \ll 1$ we can approximate the
mass spectrum as
\begin{align}
\mst{1}^2 \approx M^2 + \frac{\kLL v^2}{2}
\qqqquad 
\mst{2}^2 \approx M^2 + \frac{\kRR v^2}{2} 
\qqqquad 
\msb{}^2 = M^2  \; .
\end{align}
In that limit the oblique parameters have a particularly simple
analytical form,
\begin{align}
\aem T 
\approx \frac{\kLL^2 v^2}{64 \pi s_w^2 m_W^2} \; \frac{v^2}{M^2}
\approx \frac{\left(\mst{1}^2 - \msb{}^2 \right)^2}{16 \pi s_w^2 m_W^2 \msb{}^2} 
\qqquad \text{and} \qqquad 
S
\approx - \frac{\mst{1}^2 - \mst{2}^2}{12 \pi \msb{}^2}  \; .
\label{eq:oblique_partners}
\end{align}
Just as for the portal model, in the following we numerically test different
approaches to the dimension-6 operator matching for a set of
benchmark points given in Tab.~\ref{tab:partners-benchmarks}. Unlike
for the Higgs portal model, there is no logarithmic term $\log \Lambda/\mu$ in the scalar top partner model at one-loop level. The relevant matching schemes are:

\begin{itemize}
\item SP1: \textit{default matching,} in which the full model is
  matched to the dimension-6 effective Lagrangian in the unbroken
  phase, and assuming degenerate heavy masses $M
  \equiv \MLeft = \MRight$~\cite{hlm}. As an example,
  we show the Wilson coefficient contributing to the
  $T$ parameter,
\begin{align}
\frac{\wilson_T}{\Lambda^2} 
= \frac{1}{4(4\pi)^2\,M^2}\,
  \left[\kLL^2 - \frac{\kLR^2\,\kLL}{2\,M^2} + \frac{\kLR^4}{10\,M^4}\right] \; .
\label{eq:ct-partners-default}
\end{align} 
  The approximate form for $T$ shown in Eq.\eqref{eq:oblique_partners}
  arises already from the first term or in the limit $\kLR \to 0$.

\item SP2: \textit{non-degenerate masses,} where the heavy fields are
  integrated out following Ref.~\cite{Drozd:2015kva,Drozd:2015rsp},
  allowing for two different mass scales $\MLeft \neq
  \MRight$. For $c_T$ we find
\begin{align}
\frac{\wilson_T}{\Lambda^2}= 
\frac{\kLL^2}{4 (4\pi)^2 \MLeft^2}
& -\frac{\kLR^2 \kLL}{2 (4\pi)^2} \left[ \frac{-5\, \MLeft^2 \MRight^2+\MLeft^4 -2 \MRight^4}
                                            {2 \MLeft^2 ( \MLeft^2-\MRight^2 )^3}
  +\frac{3 \, \MRight^4}{( \MLeft^2-\MRight^2 )^4} \, \log \frac{\MLeft^2}{\MRight^2} \right] \notag \\
& +\frac{\kLR^4}{2 (4\pi)^2} \left[ \frac{10 \MLeft^2 \MRight^2+\MLeft^4+\MRight^4}
                                       {2 \MLeft^2 (\MLeft^2-\MRight^2)^4}
  +\frac{3 \MRight^2 (\MLeft^2+ \MRight^2)}{(\MRight^2-\MLeft^2 )^5} \, \log \frac{\MLeft^2}{\MRight^2} \right] \; .
\label{eq:ct-partners-nondeg}
\end{align}  
  This form reduces to
  Eq.\eqref{eq:ct-partners-default} in the limit of one mass scale
  only.

\item SP$v$: \textit{$v$-improved matching,} where the two heavy
  scales $\MLeft$ and $\MRight$ are traded for $\mst{1}\approx \msb{}$
  and $\mst{2}$, respectively. We then find
\begin{align}
\frac{\wilson_T}{\Lambda^2}
= \frac{\kLL^2}{4 (4\pi)^2 \mst{1}^2}
&-\frac{\kLR^2 \kLL}{2 (4\pi)^2} \left[ \frac{-5 \mst{1}^2 \mst{2}^2+ \mst{1}^4 -2 \mst{2}^4}
                                           {2 \mst{1}^2 (\mst{1}^2-\mst{2}^2 )^3}
 +\frac{3 \mst{2}^4}{( \mst{1}^2-\mst{2}^2 )^4} \, \log \frac{\mst{1}^2}{\mst{2}^2} \right] \notag \\
&+\frac{\kLR^4}{2 (4\pi)^2} \left[ \frac{10 \mst{1}^2 \mst{2}^2+ \mst{1}^4 + \mst{2}^4}
                                      {2 \mst{1}^2 (\mst{1}^2-\mst{2}^2 )^4}
 +\frac{3 \mst{2}^2 (\mst{1}^2+\mst{2}^2)}{(\mst{2}^2-\mst{1}^2 )^5} \, \log \frac{\mst{1}^2}{\mst{2}^2} \right] \; .
\label{eq:ct-partners-improved}
\end{align} 
  which has
  same functional for as in Eq.\eqref{eq:ct-partners-nondeg}.
  In this expression the two left-handed masses $\mst{1}$ and $\msb{}$
  can be used interchangeably. 

\item BP1: \textit{broken-phase matching,} in which case the Wilson
  coefficients are derived through explicit matching in the broken
  phase, assuming a single degenerate heavy scale $M$. We here obtain
\begin{align}
\frac{\wilson_T}{\Lambda^2} 
= \frac{1}{640(4\pi)^2\,M^2}\,
& \left\{ 5 (19 \kLL^2 + 10 \kLL\kRR + 3 \kRR^2) - 
         15 (3 \kLL + \kRR) \frac{\kLR^2}{M^2} + 
         8 \frac{\kLR^4}{M^4} \right. \notag \\ 
&+ \left. \left[ 20 (\kLL - \kRR) (3 \kLL + \kRR) + 
         5 (\kLL + \kRR) \frac{\kLR^2}{M^2} - 9 \frac{\kLR^4}{M^4} \right] \cft \right. \notag \\      
& \left. + 
   5 \left[ (\kLL - \kRR)^2 + (5 \kLL - \kRR) \frac{\kLR^2}{M^2} \right] \ceit + 
            \frac{\kLR^4}{M^4} c_{12\tilde{t}} \right\} \; .
\label{eq:ct-partners-explicit}
\end{align}
  As in the expression for the masses in
  Eq.\eqref{eq:spartner-masses-redefined} the appearance of the mixing
  angles leads to an additional implicit dependence of on mass matrix parameter
  $M$ and $\kappa_j$, which we keep in the interest of a compact
  formula.  However, this additional dependence obscures the link to
  the simple form of $\wilson_T$ in Eq.\eqref{eq:ct-partners-default}.
  The new terms appearing through broken-phase matching, as compared
  to Eq.\eqref{eq:ct-partners-default} are proportional to $\st$ and
  hence suppressed for weakly-coupled scenarios.  This is different
  from the Higgs portal case, where broken-phase matching captures
  finite terms without an extra suppression at small mixing,
  Eq.~\eqref{eq:singlet-explicit}.  This difference can again be
  explained by the fact that scalar partner effects occur though
  loops, whereas for the Higgs portal both tree-level and loop-induced
  operators co-exist.

\item BP$v$: \textit{$v$-improved broken-phase matching,} where two
  separate heavy scales $\mst{1}$ and $\mst{2}$ are included. For
  $\wilson_T$, we compute the 1PI two-point Greens function
  combination $\Pi_{WW}(0)-\cwd\,\Pi_{ZZ}(0)$ in the full model,
  expand in powers of $v/\MLeft$ and $v/\MRight$ separately, and match
  to the dimension-6 effective Lagrangian result. Then we replace
  $\MLeft$ and $\MRight$ by $\mst{1}$ and $\mst{2}$ and introduce
  $\kLLtilde = \ctd\kLL + \std\kRR$ and $\kRRtilde = \std
  \kLL + \ctd\kRR$.  The result is lengthy, but we can illustrate its
  main features by retaining the leading dependence on the splitting
  $(\mst{2}^2 - \mst{1}^2)/\mst{2}^2$,
   \begin{align}
\frac{\wilson_T}{\Lambda^2} \supset 
\frac{(\mst{2}^2-\mst{1}^2)}{2560\,(4\pi)^2\,\mst{2}^4\,v^4}\,&\Bigg{\{}
56 \kLR^4 - \kLR^2 \mst{2}^2\,\left[240 \kLLtilde + 48\kRRtilde \right] + \mst{2}^4\left[295 \kLLtilde^2+90\kLLtilde\kRRtilde + 15\kRRtilde^2 \right] \notag \\
&  -4 \left[12 \kLR^4 - \kLR^2 (5 \kLLtilde + 11 \kRRtilde) \mst{2}^2+ 5 (-15 \kLLtilde^2 + 2 \kLLtilde \kRRtilde + \kRRtilde^2) \mst{2}^4\right] c_{4\tilde{t}} \notag \\
&+ \left[-24 \kLR^4 + 16 \kLR^2 (13 \kLLtilde + \kRRtilde) \mst{2}^2+ 5 (9 \kLLtilde^2 - 10 \kLLtilde \kRRtilde + \kRRtilde^2) \mst{2}^4) c_{8\tilde{t}}\right] \notag \\
&+ \kLR^2\,\left[16 \kLR^2 +12(\kLLtilde - \kRRtilde)\,\mst{2}^2 \right] c_{12\tilde{t}}
\Bigg{\}}
%\label{eq:ct-partners-nondeg}
\; ,
  \end{align}

\item BP$v$': The definition of BP$v$ above is based on the assumption
  that $\st$ is small. If that is not the case, one arrives at a more
  accurate result by performing the expansion of the full model
  directly in terms of $v/\mst{1}$ and $v/\mst{2}$, while keeping the
  power counting $\st \sim \ord(v/M)$ and $\msb{}-\mst{1} \sim
  \ord(v)$. One then finds
\begin{align}
\frac{\wilson_T}{\Lambda^2} =
 \frac{1}{16\pi^2 v^4} \biggl [ 3s_{\tilde{t}}^4 \biggl (
 &\mst{1}^2 + \mst{2}^2 + \frac{2\mst{1}^2 \mst{2}^2}{\mst{2}^2-\mst{1}^2}\,
 \log\frac{\mst{1}^2}{\mst{2}^2} \biggr ) \notag \\ &+
 \frac{3s_{\tilde{t}}^2(\msb{}^2-\mst{1}^2)}{\mst{2}^2-\mst{1}^2} \biggl (
  3\mst{2}^2 - \mst{1}^2 - \frac{2 \mst{2}^2}{\mst{2}^2-\mst{1}^2}\,
  \log\frac{\mst{1}^2}{\mst{2}^2}
\biggr ) + \frac{(\msb{}^2-\mst{1}^2)^2}{\mst{1}^2} \biggr ]
\label{eq:ct-partners-bpvp}
\end{align}
  Note that Eq.\eqref{eq:ct-partners-bpvp} is consistently of order
  $\ord(v^2/M^2)$ as required by the EFT approach.
\end{itemize}
\bigskip

%-----------------------------------------------------
\begin{table}[t]
\begin{tabular}{l|rrrrr|rrrr|rrrr}
\hline   
& $\MLeft$ & $\MRight$& $\kLL$ & $\kRR$ & $\kLR$ & $\mst{1}$ & $\mst{2}$ & $\msb{}$ & $\theta_{\tilde{t}}$ & \multicolumn{4}{c}{$ c_T v^2/\Lambda^2$}\\ 
&&&&&&&&&& SP1 & SP$v$ & BP$v$  & BP$v$' \\ \hline
  P1 & 500 & 500 & -0.34 & 0.00 & 0.58   & 490 & 500 & 500 & -0.01 & 4.23$\times 10^{-5}$& 4.41$\times 10^{-5}$ & 4.40$\times 10^{-5}$ & $4.40 \times 10^{-5}$ \\ 
  P2 & 500 & 500 & 0.066 & 2.89 &   74.5 & 500 & 580 & 500 & -0.15 & 1.35$\times 10^{-6}$ & 5.69$\times 10^{-6}$& 2.76$\times 10^{-6}$ & $1.38 \times 10^{-6}$ \\         
  P3 & 490 & 500 & 0.1 & 0.1 & 0.1 & 493 & 503 & 490 & -0.0017 & 3.71$\times 10^{-6}$ & 3.82$\times 10^{-6}$ & 3.82$\times 10^{-6}$ & $3.82 \times 10^{-6}$ \\
  P4 & 450 & 500 & 0.1 & 0.1 & 0.1 & 453 & 503 & 450 & -0.00036 & 3.71$\times 10^{-6}$& 4.52$\times 10^{-6}$ & 4.52$\times 10^{-6}$ & $4.52 \times 10^{-6}$ \\
\hline
\end{tabular}    
\caption{Benchmark points for the scalar partner model, where all masses are
  given in GeV.}
\label{tab:partners-benchmarks}
\end{table}
%-----------------------------------------------------

For our numerical analysis, we again define some
benchmark points in Tab.~\ref{tab:partners-benchmarks}. For P1 we
assume a single heavy mass scale $M$ and small mixing angle
$\theta_{\tilde{t}}$, leading to mild mass splittings between the
physical heavy-quark partners; for P2 we also use a single heavy mass
scale $M$ but a larger mixing angle from a stronger coupling to the
Higgs sector; P3 and P4 both have non-degenerate top partners driven by non-degenerate heavy mass scales in the unbroken
phase, $\MLeft \neq \MRight$.  For these two scenarios the mixing angle is
tiny.\bigskip

% -----------------------------------
\begin{table}[b!]
 \begin{tabular}{l|l|rrrrrrr} \hline
&  & full model & SP1 & SP2 & SP$v$ & BP1 & BP$v$ & BP$v'$ \\ \hline   
\multirow{4}{*}{$S$} & P1 & $ 0.11\times 10^{-2}$ & $ 0.11\times 10^{-2}$ & $ 0.11\times 10^{-2}$ & $ 0.11\times 10^{-2}$ & $ 0.11\times 10^{-2}$ & $ 0.11\times 10^{-2}$ & $0.11\times 10^{-2}$ \\
                     & P2 & $-0.15\times 10^{-3}$ & $-0.13\times 10^{-3}$ & $-0.13\times 10^{-3}$ & $-0.34\times 10^{-3}$ & $-0.58\times 10^{-3}$ & $-0.60\times 10^{-3}$ & $-0.15\times 10^{-3}$ \\
                     & P3 & $-0.32\times 10^{-3}$ & $-0.31\times 10^{-3}$ & $-0.31\times 10^{-3}$ & $-0.32\times 10^{-3}$ & $-0.31\times 10^{-3}$ & $-0.32\times 10^{-3}$ & $-0.32\times 10^{-3}$ \\
                     & P4 & $-0.38\times 10^{-3}$ & $-0.31\times 10^{-3}$ & $-0.38\times 10^{-3}$ & $-0.38\times 10^{-3}$ & $-0.31\times 10^{-3}$ & $-0.38\times 10^{-3}$ & $-0.38\times 10^{-3}$ \\ \hline 
\multirow{4}{*}{$T$} & P1 & $ 0.55\times 10^{-2}$ & $ 0.54\times 10^{-2}$ & $ 0.54\times 10^{-2}$ & $ 0.56\times 10^{-2}$ & $ 0.54\times 10^{-2}$ & $ 0.56\times 10^{-2}$ & $0.56\times 10^{-2}$ \\
                     & P2 & $ 0.18\times 10^{-3}$ & $ 0.17\times 10^{-3}$ & $ 0.17\times 10^{-3}$ & $ 0.78\times 10^{-3}$ & $ 0.17\times 10^{-3}$ & $ 0.35\times 10^{-2}$ & $ 0.18\times 10^{-3}$ \\ 
                     & P3 & $ 0.49\times 10^{-3}$ & $ 0.48\times 10^{-3}$ & $ 0.48\times 10^{-3}$ & $ 0.49\times 10^{-3}$ & $ 0.48\times 10^{-3}$ & $ 0.49\times 10^{-3}$ & $0.49\times 10^{-3}$ \\
                     & P4 & $ 0.58\times 10^{-3}$ & $ 0.48\times 10^{-3}$ & $ 0.48\times 10^{-3}$ & $ 0.58\times 10^{-3}$ & $ 0.48\times 10^{-3}$ & $ 0.58\times 10^{-3}$ & $0.58\times 10^{-3}$ \\\hline
  \end{tabular}
 \caption{Predictions for $S$ and $T$ in the scalar partner extension
   for the full model and the different effective Lagrangian
   setups. The benchmark points are defined in
   Tab.~\ref{tab:partners-benchmarks}.}
 \label{tab:spartners-stu}
\end{table}
% -----------------------------------

In Tab.~\ref{tab:spartners-stu} we show the $S$ and $T$ parameters for
the different matching schemes and each of the benchmark points.  In
general, the effective Lagrangian succeeds in reproducing the full
results when the new physics is weakly coupled and relatively heavy,
\eg for P1. Small mass splittings $\mst{1}^2-\mst{2}^2 \ll \MRight^2$
are compatible with the assumption of a single heavy scale in the
default matching setup. In this weakly-coupled, small mixing regime
the $v$-improved corrections have a tiny numerical impact. A challenge
for the effective Lagrangian is a larger mass splitting. This occurs
in the unbroken phase when $\MLeft \neq \MRight$. In this case,
illustrated by P3 and P4, $v$-improvement corrects for these
$\ord(\mst{1}^2-\mst{2}^2)$ key effects in the oblique
parameters~\eqref{eq:oblique_partners}. On the other hand, the use of
$v$-improvement becomes less straightforward for the case of large
mixing, as illustrated by P2. The SP$v$ and BP$v$ schemes, which are
based on the assumption that the $\tilde{1}$ and $\tilde{2}$ are
closely aligned with the $\tilde{Q}$ and $\tilde{t}_R$ states, differ
drastically from the full model.  On the other hand, the BP$v'$
scheme, based on a direct expansion in $1/\mst{1,2}$ reproduces all
benchmark scenarios very well. Thus it is important to note that the
principle of $v$-improved matching is not uniquely defined, but its
optimal implementation needs to be worked out separately for each
specific model.

%-----------------------------------------------------
\begin{figure}[t]
\includegraphics[width=0.4\textwidth]{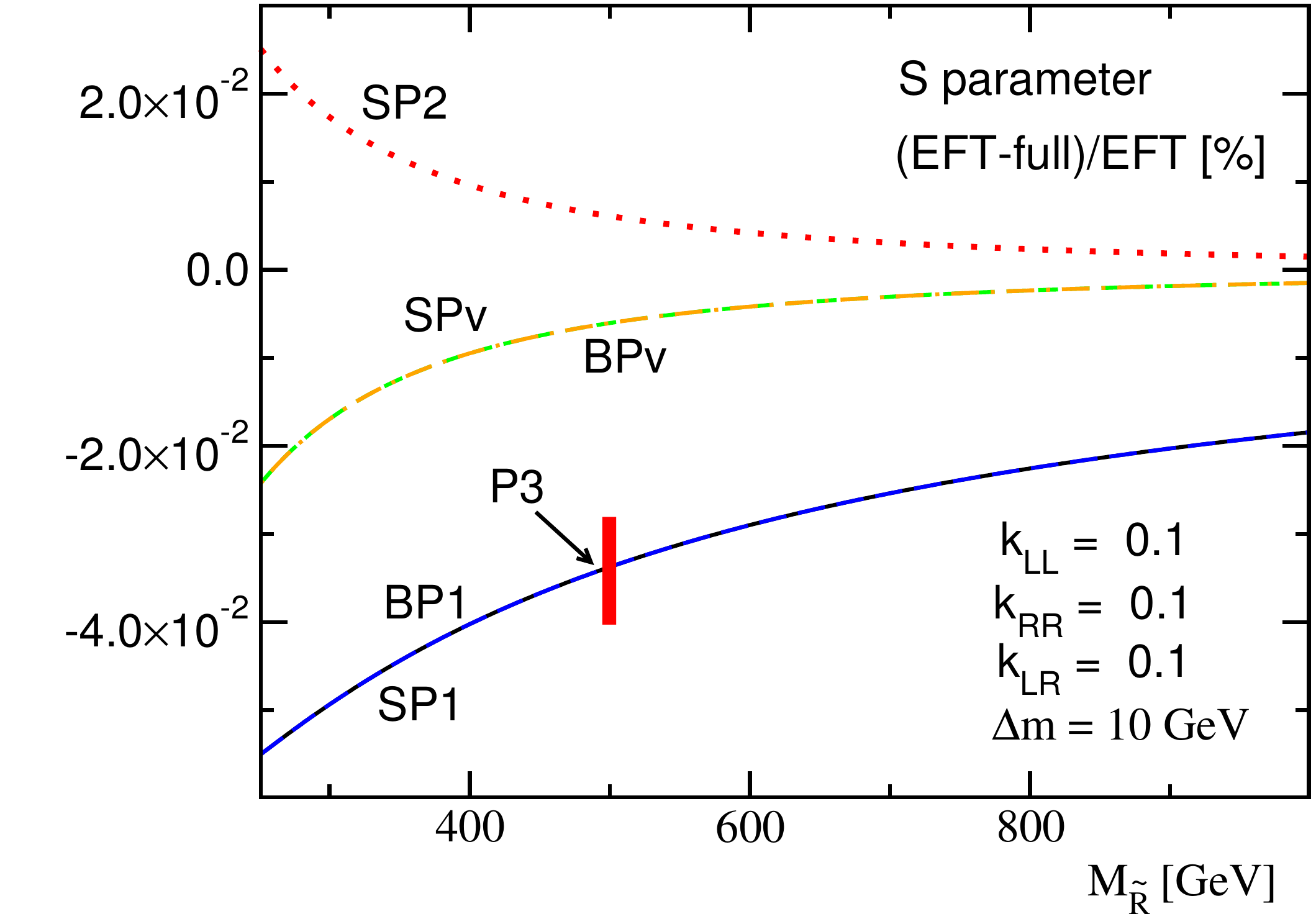}
\hspace*{0.08\textwidth}
\includegraphics[width=0.4\textwidth]{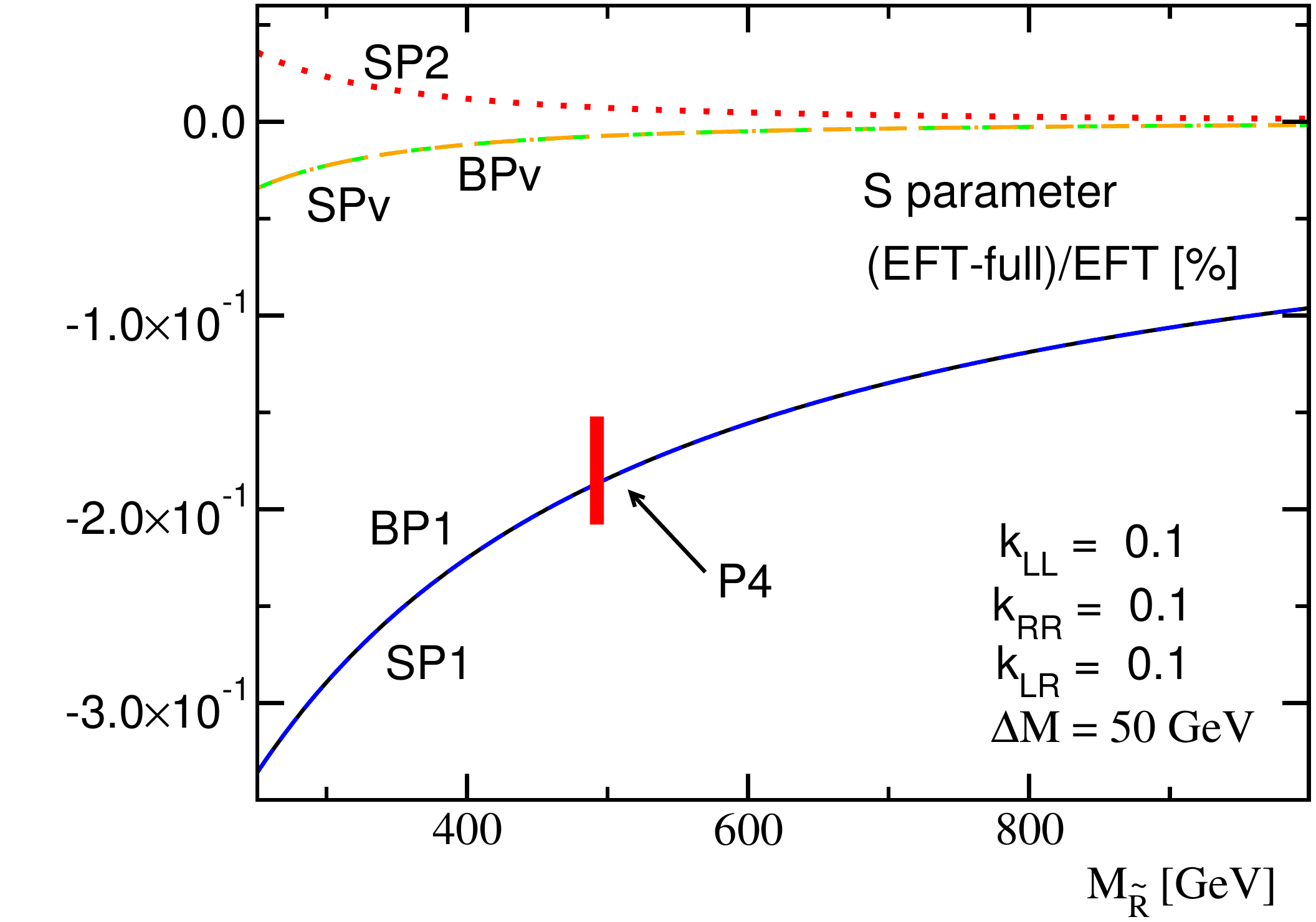} \\
\includegraphics[width=0.4\textwidth]{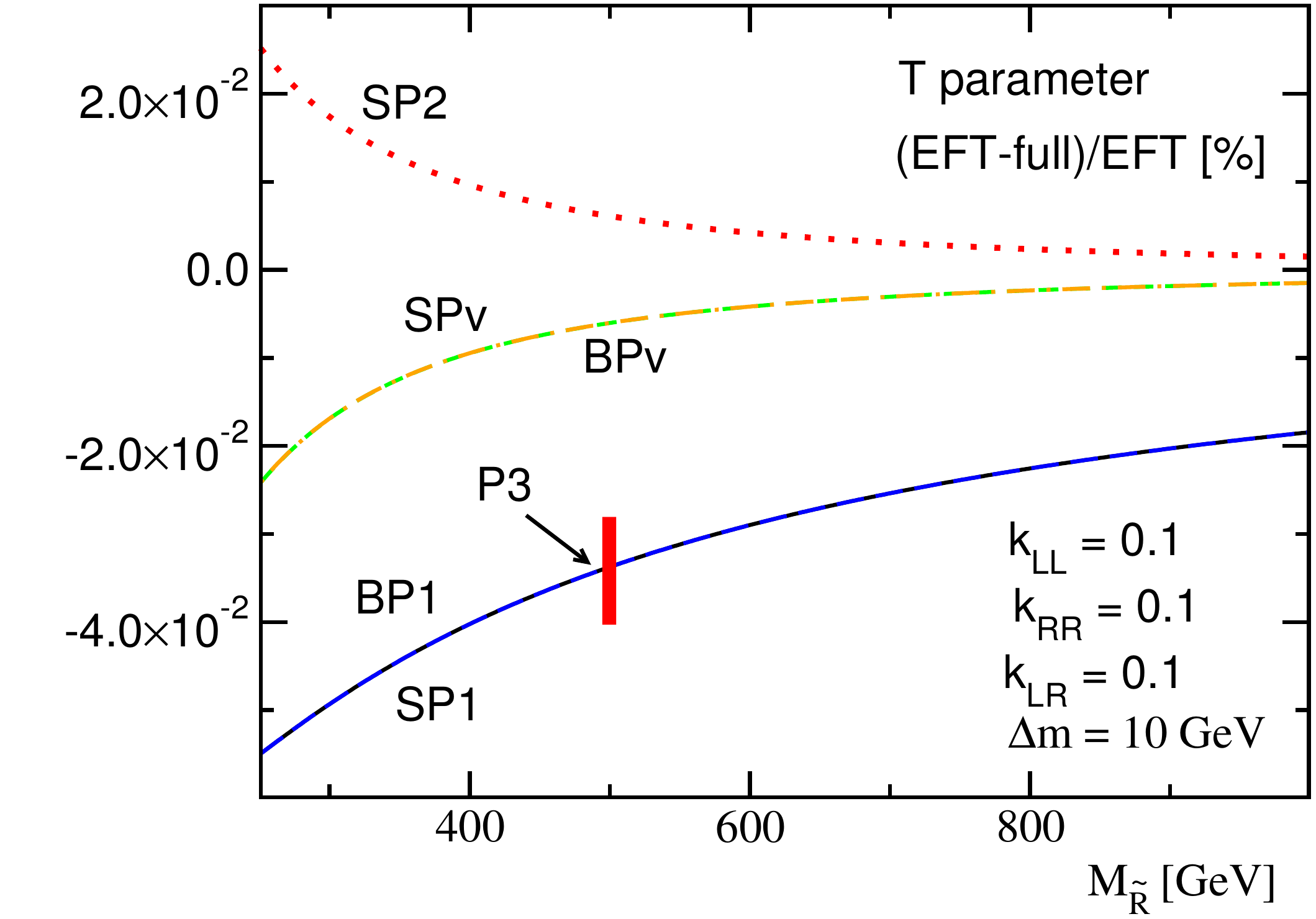}
\hspace*{0.08\textwidth}
\includegraphics[width=0.4\textwidth]{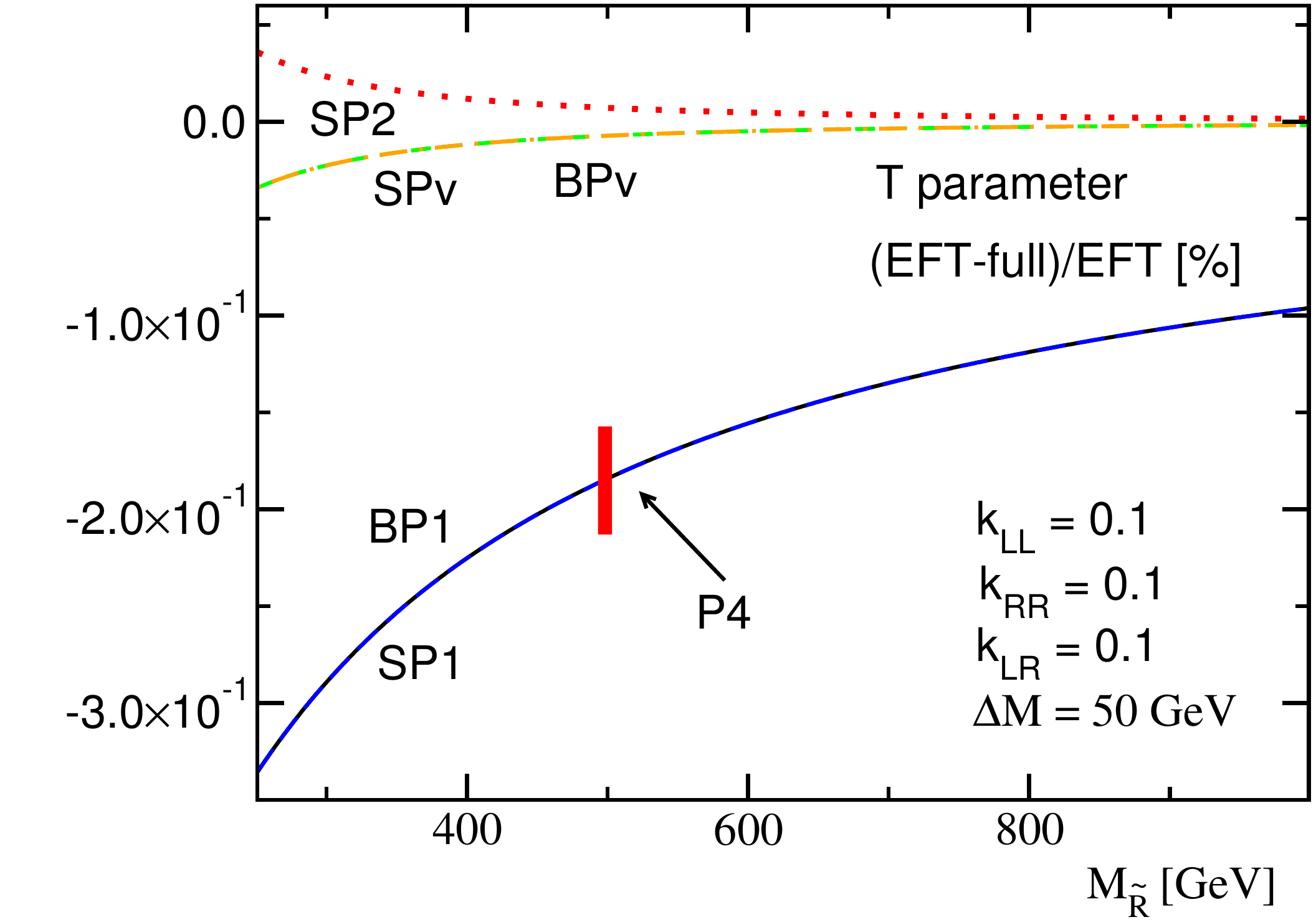} 
\caption{Deviation of $S$ and $T$ from the full model prediction as a
  function of $\MRight$. We consider two mass splittings $\Delta M
  \equiv \MRight- \MLeft = 10$~GeV (left) and 50~GeV (right). The red
  marks indicate the benchmark points P3 and P4 from
  Tab.~\ref{tab:partners-benchmarks}.}
\label{fig:ST-partners-decoupling}
\end{figure}
%-----------------------------------------------------

In Fig.~\ref{fig:ST-partners-decoupling} we see that all matching
schemes essentially follow the expected decoupling behavior at the few
per-cent level.  Differences appear from the way the different schemes
account for the non-degenerate top partner masses.  Enhancing the mass
splitting leads to significantly larger deviations from the full model
for effective Lagrangian setups where only one single mass scale is
assumed. 

Furthermore, we observe a remarkable contribution from the
vev-dependent contributions of dimension greater than six, included
via $v$-improvement; for our parameter choices they flip the sign of
the deviations from the full model.  Such a systematic positive
(negative) offset can be understood by the comparably weaker
suppression of the $v$-improved Wilson coefficients, which scale as
inverse powers of the physical masses $\wilson_i \sim 1/\mst{i}^2$,
one of them being lighter than the intrinsic heavy mass scales.  A
similar trend is observed for the scalar singlet effective Lagrangian
in Fig.~\ref{fig:ST-singlet-decoupling}, although the behavior there
is affected by logarithmic modulations and sensitive to the additional
$v$-improved replacement $\lambda_3^2/(2\lambda_2) \to
2(1-\cos\alpha)$.
 
The additional improvement from broken phase matching is barely
visible for each of the scenarios. This is due to the fact that in the
scalar top model there are no contributions with both heavy particles
and SM particles in the loop, in contrast to the Higgs singlet
model. 
%Also for the mixed heavy-light loops, the broken-phase matching leads to
%numerically relevant differences.
For the mixed heavy-light loops, the broken-phase matching
leads to important differences due to the non-negligible mass of the SM
gauge and Higgs bosons.

%-----------------------------------------------------
\begin{figure}[t]
\includegraphics[width=0.4\textwidth]{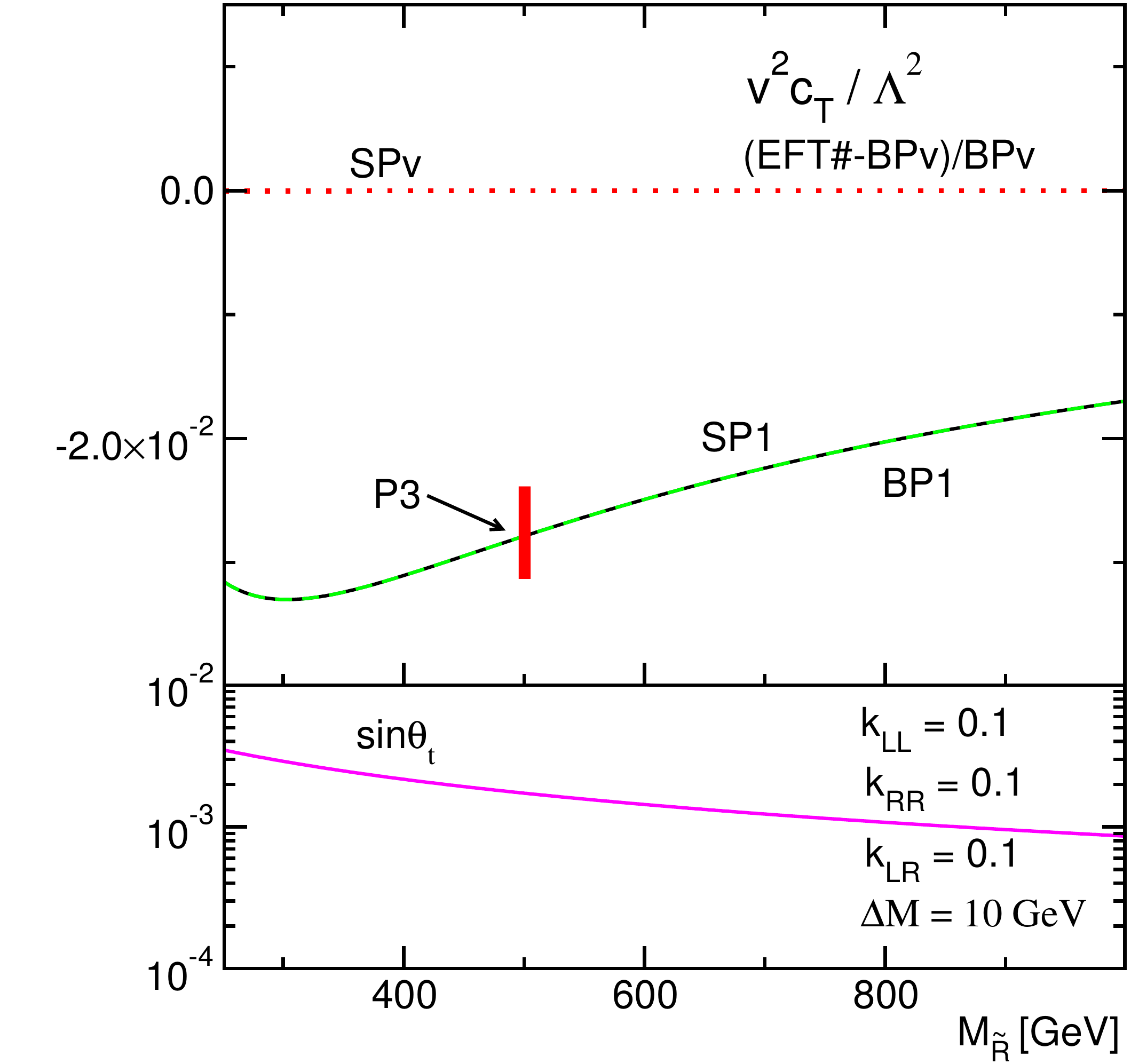}
\hspace*{0.08\textwidth}
\includegraphics[width=0.4\textwidth]{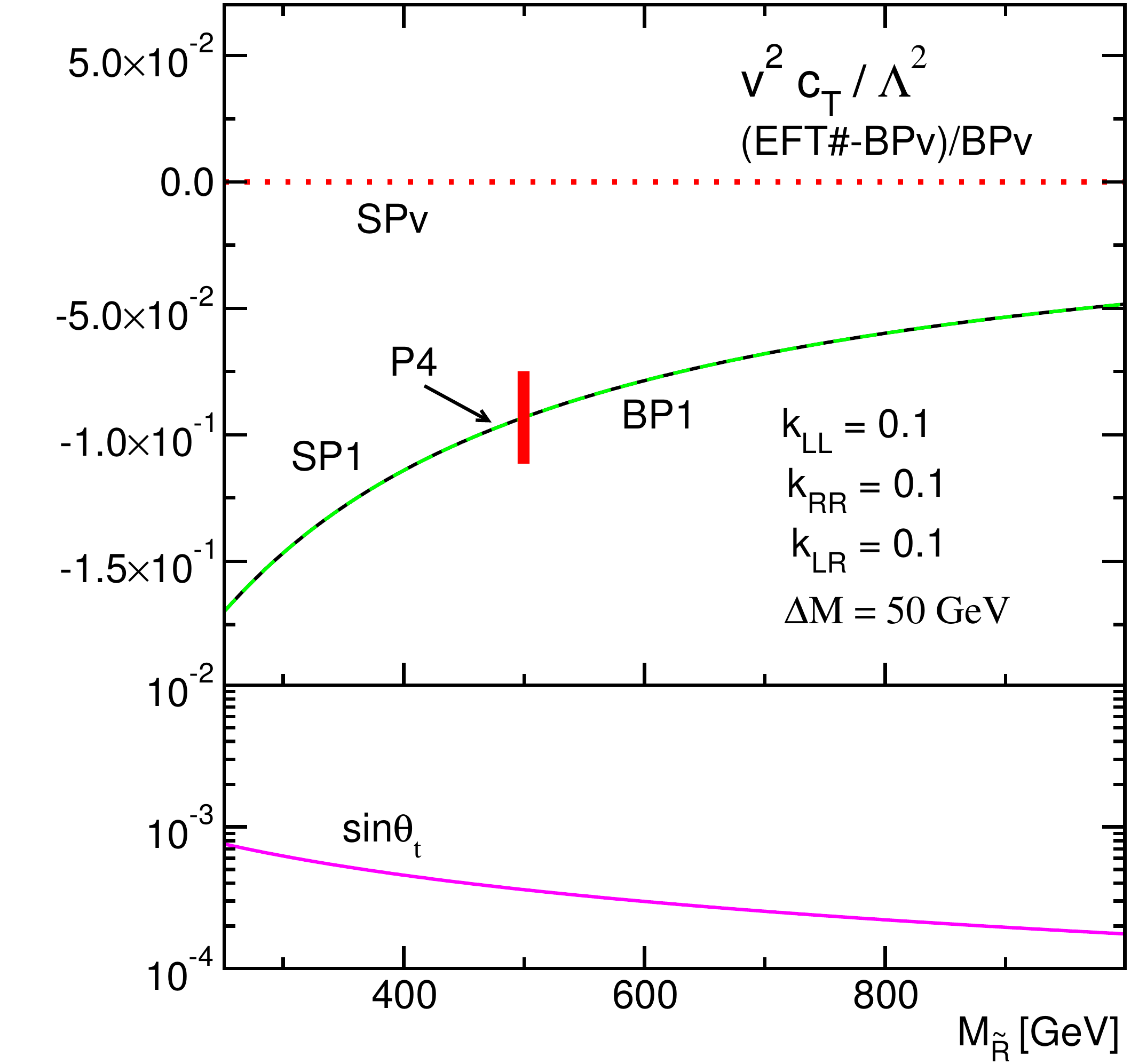} 
\caption{Wilson coefficient $\wilson_T$ as a function of the heavy
  partner mass $\MRight$. The different curves show the relative
  deviation between the different setups relative to 
  BP$v$. We consider two mass splittings $\Delta M \equiv \MRight-
  \MLeft = 10$~GeV (left) and 50~GeV (right). The red marks indicate
  the benchmark points P3 and P4 from
  Tab.~\ref{tab:partners-benchmarks}. The decoupling behavior of the
  stop mixing angle is shown in the lower sub-panels.}
\label{fig:ST-wilson-partners-decoupling}
\end{figure}
%-----------------------------------------------------

As complementary information we show the $\MRight$-dependence of the
Wilson coefficients $\wilson_T$ in
Fig.~\ref{fig:ST-wilson-partners-decoupling}.  We now compare the
different matching schemes to the BP$v$ choice, which
includes broken-phase matching, $v$-improvement, and non-degenerate
heavy mass scales. The lower panels illustrate the consistent
evolution of the top partner mixing angle towards the decoupling
limit. Again, for matching prescriptions assuming a single heavy scale
we observe deviations rapidly increasing with the scale separation
$\Delta M$.\bigskip

%-----------------------------------------------------
\begin{figure}[b!]
\includegraphics[width=0.33\textwidth]{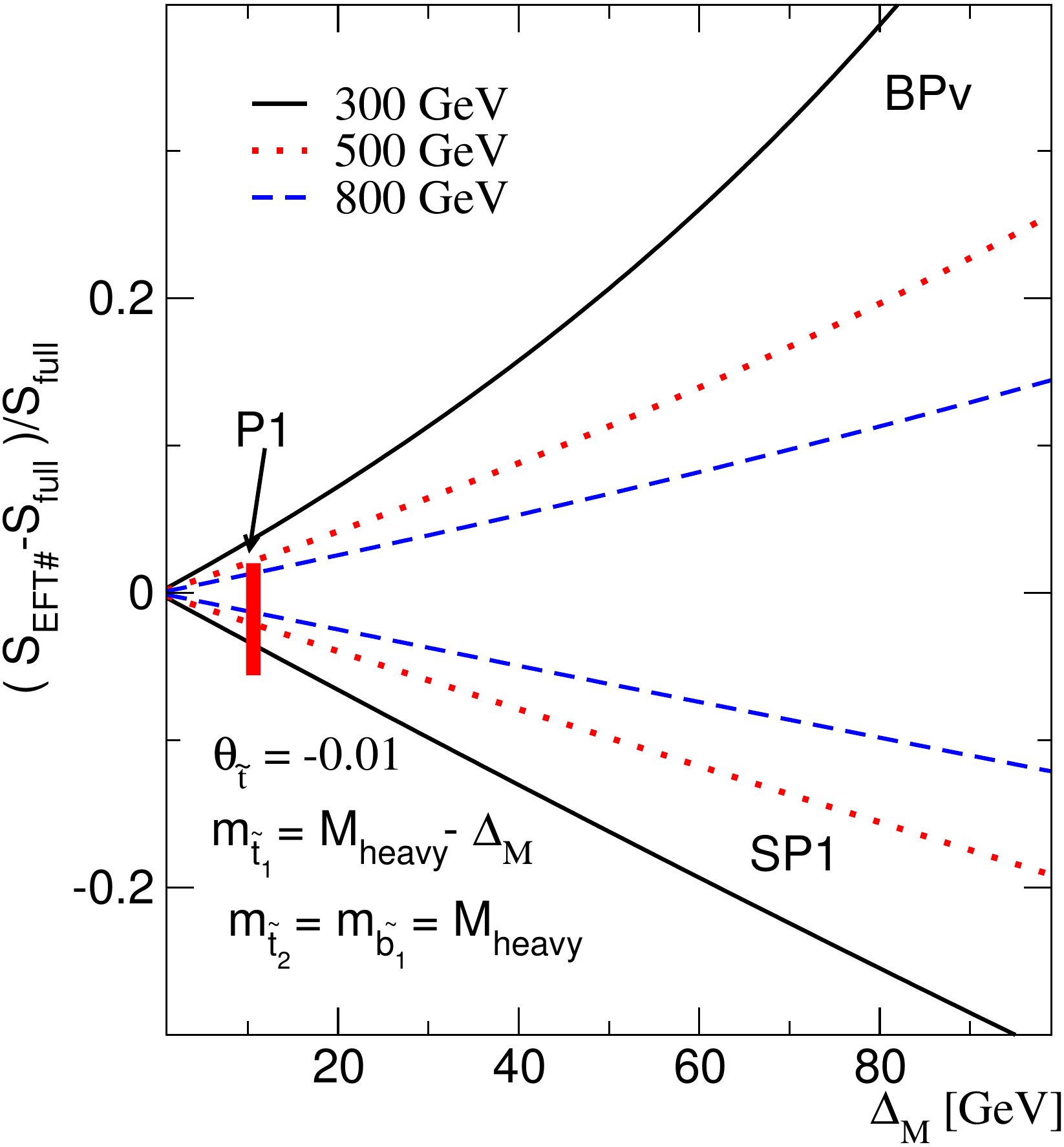}
\includegraphics[width=0.33\textwidth]{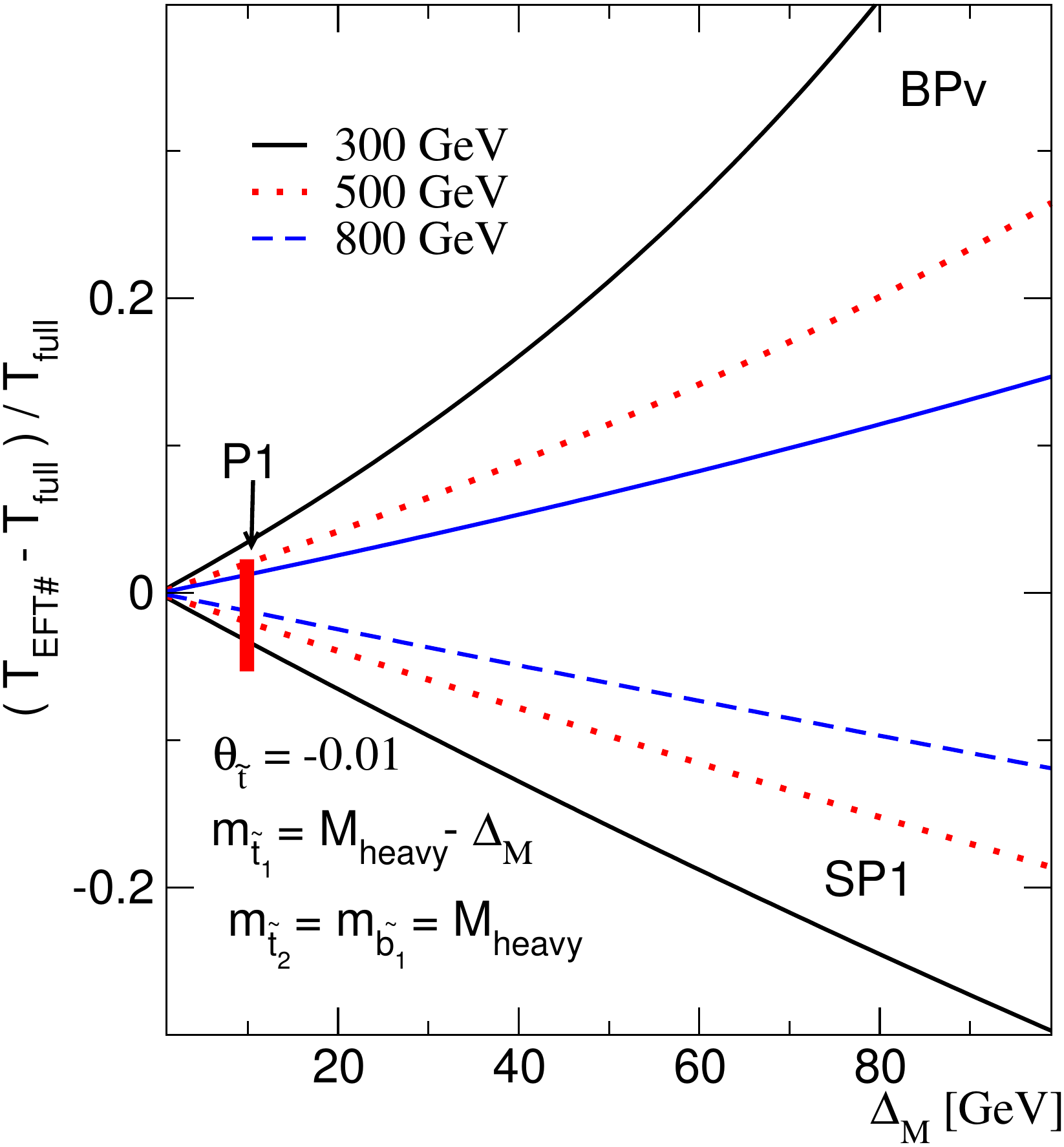}
\includegraphics[width=0.295\textwidth]{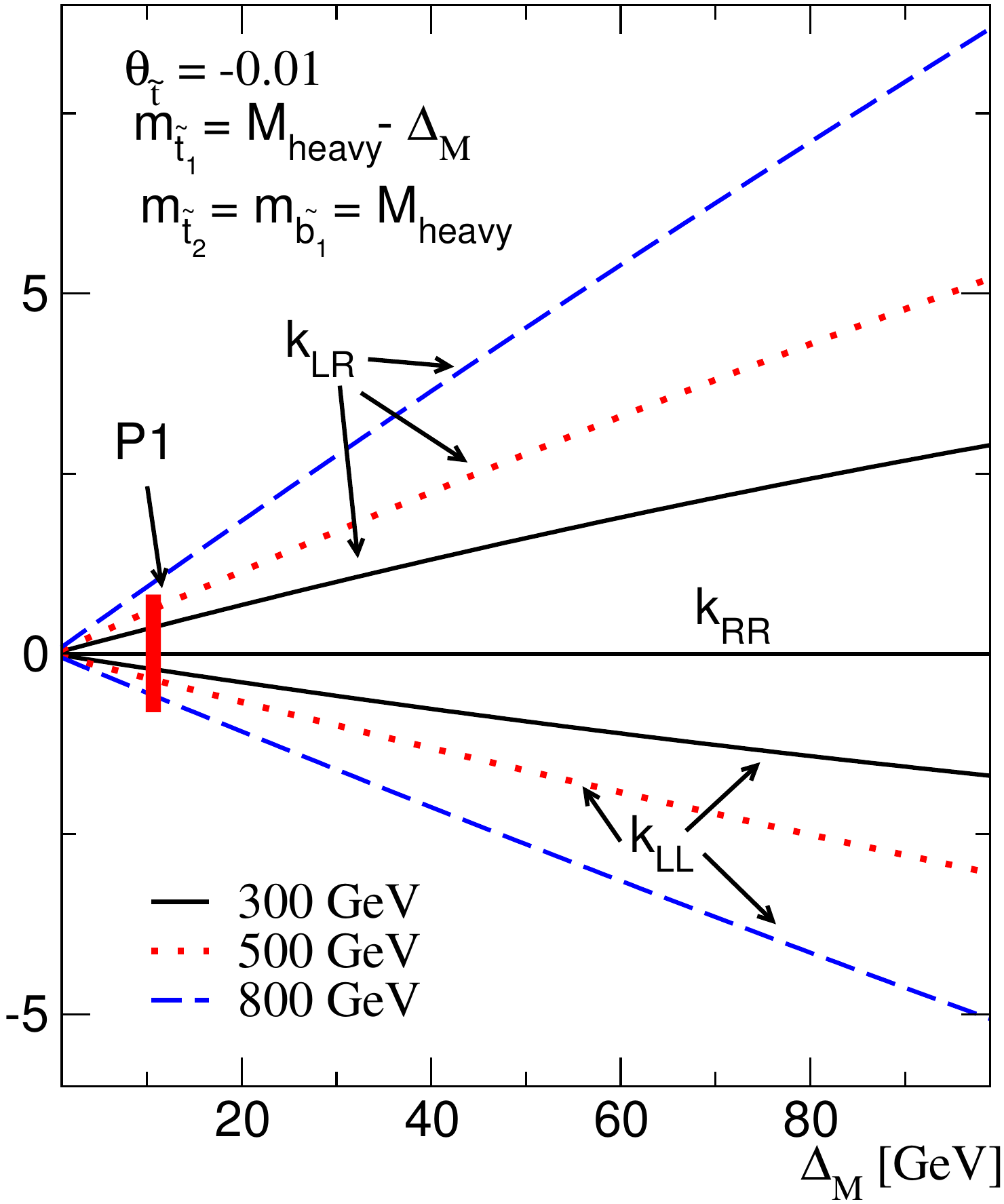}
\caption{Relative difference $(S_\text{full}-
  S_\text{EFT})/S_\text{full}$ (left) and $(T_\text{full}-
  T_\text{EFT})/T_\text{full}$ (center) as a function of the mass
  splitting between the scalar partners. The mixing angle is fixed to
  $\theta_{\tilde{t}} = -0.01$.  The right panel shows the couplings
  to the Higgs sector, with $\kLR$ given in units of GeV. The red bar
  locates the benchmark point P1 from
  Tab.~\ref{tab:partners-benchmarks}.}
\label{fig:overmsplitting}
\end{figure}
%-----------------------------------------------------

As alluded to above, large mass splittings are the most serious
obstacle in constructing an accurate effective description of the
heavy scalar partner sector, in particular when they emerge in the broken
phase via large couplings $\kappa_i$.  For
Fig.~\ref{fig:overmsplitting} we introduce a mass splitting in the
broken phase with $\mst{1} + \Delta M= \mst{2} = \msb{} = M$, where
$M$ sets a common mass scale of the heavy gauge eigenstates $\MLeft =
\MRight =M$, and $\Delta M$ is generated primarily through the
$v$-induced $\kLL$ term in Eq.\eqref{eq:spartner-masses-redefined}.
We consider three different values for $M$ and a mixing angle
$\theta_{\tilde{t}} = -0.01$. For each value of $\Delta M$ we show the
relative deviation of the effective Lagrangian prediction from the
full model in the left and center panels.  The right panel shows the
corresponding variation of the couplings $\kappa_j$. In general,
decoupling leads to decreasing discrepancies between the full model
and the effective theory.  With increased Higgs couplings, the mass
splittings and thus the deviations increase. Interestingly, the
observed departures show some systematic behavior, with default
matching underestimating and $v$-improved matching overestimating the
full model results.  This can be understood as follows: in the limit
of small mixing $\st$ and small mass differences $\Delta M$, the
results for the $T$ parameter in full model, the default matching
(SP1), and the $v$-improved broken-phase matching (BP$v$) can be
approximated as
\begin{align}
\aem T_{\text{full}} &= \frac{(\msttwo^2 - \mstone^2)^2}{8\pi\swd\,m_W^2\,(\mstone^2+\msttwo^2)}
 + \ord\bigl(\st,(\Delta M)^4\bigr) \notag \\
\aem T_{\text{SP1}}
&= \frac{(\msttwo^2 - \mstone^2)^2}{16\pi\swd\,m_W^2\,\msttwo^2}
 + \ord\bigl(\st,(\Delta M)^4\bigr) \notag \\
%\label{eq:SP1} 
\aem T_{\text{BP}v}
&= \frac{(\msttwo^2 - \mstone^2)^2}{16\pi\swd\,m_W^2\,\mstone^2}
 + \ord\bigl(\st,(\Delta M)^4\bigr) \, .
\label{eq:BPv}
 \end{align}
Due to the mass splitting $\msttwo - \mstone > 0$, one thus obtains
$T_{\text{SP1}} < T_{\text{full}} < T_{\text{BP}v}$ in the scenario in
Fig.~\ref{fig:overmsplitting}. Similar relations hold for the $S$
parameter.

Thus both SP1 and BP$v$ deviate from the full model due to $v$-induced
effects, which in this case are not captured by the $v$-improvement,
but are intrinsic to the EFT expansion itself.

%%%%%%%%%%%%%%%%%%%%%%%%%%%%%%%%%%%%%%%%%%%%%%%%%%%%%%%%%%%%
\section{Higgs decay to photons}
\label{sec:photons}

While LHC observables in general are dominated by tree-level effects
from new physics, there are a few select operators where loop-level
modifications can make a difference. These include the Wilson
coefficients $\wilson_\gamma$ and $\wilson_g$, since also in the
Standard Model they are generated only at the loop level. This is why
in many parametrizations these Wilson coefficients are scaled
differently~\cite{legacy}. As part of our analysis
of loop effects, we include contributions to the decay $h\to
\gamma\gamma$ of the SM-like Higgs from electrically charged heavy
states in the loop. For example, a charged scalar can be added to the
Standard Model contributions, which stem from heavy fermions and the $W$-boson,
\begin{align}
 \lag \supset -\frac{g_{h\gamma\gamma}}{4v} \,F^{\mu\nu}\,F_{\mu\nu}\,\PHiggs 
 \quad \text{with} \quad
  g_{h\gamma\gamma} &= 
  g_{h\gamma\gamma}^\text{SM} -\frac{\aem}{\pi}\,
  C_S Q_S^2 \,\frac{g_{hSS} v}{2m_S^2}\, A_S(\tau_S) \notag \\
&= -\frac{\aem}{\pi}\,
 \left[ \sum_{f = t,b,\tau} \, C_f Q^2_f\, A_f(\tau_f) 
      + A_W \,(\tau_W) 
      + C_S Q_S^2 \,\frac{g_{hSS} v}{2m_S^2}\, A_S(\tau_S)
\right] \, .
 \label{eq:haalag}
\end{align}
Here, $F_{\mu \nu}$ is the abelian photon field strength, and $C_f$
and $C_S$ include potential color factors, distinguishing for example
charged Higgs bosons from scalar top partners. The $A_{f,W,S}(\tau)$
are loop functions~\cite{review}, which for on-shell
Higgs decays can be expanded in the ratio $1/\tau_j = m_h^2/(4m_j^2)$,
\begin{align}
A_f(\tau) &= 
 \frac{2}{3}  + \frac{7}{45\,\tau} + \frac{4}{63\tau^2} + \frac{52}{1575\tau^3}+  \ord \left( \frac{1}{\tau^4} \right)\,, \notag \\
A_W(\tau) &= 
 -\frac{7}{2}  - \frac{11}{15\,\tau} - \frac{38}{105\tau^2} - \frac{116}{525\tau^3} +\ord \left( \frac{1}{\tau^4} \right)\,, \notag \\
A_S(\tau) &=
 \frac{1}{6} + \frac{4}{45\,\tau} + \frac{2}{35\tau^2} + \frac{64}{1575\tau^3} + \ord \left( \frac{1}{\tau^4} \right) \; .
 \label{eq:loopfunctions}
\end{align}
As is well known, the size of the loop--induced contributions
increases with the spin of particle in the loop, and vectors
contribute with opposite sign from fermions and scalars (assuming a
positive coupling to the Higgs boson).  Finally, the $\tau$ dependence
relative to the low--energy limit $\tau \to \infty$ is different as
well,
\begin{align}
\frac{A_f(\tau)}{A_f(\infty)}
 &\approx 1 + 0.06 \; \frac{m_h^2}{m_f^2}\,,  &
\frac{A_W(\tau)}{A_W(\infty)} 
 &\approx 1 + 0.05 \; \frac{m_h^2}{m_W^2}\,, &
\frac{A_S(\tau)}{A_S(\infty)} 
 &\approx 1 + 0.13 \; \frac{m_h^2}{m_S^2}  \; ,
\label{eq:loop-relative}
\end{align}
with the largest mass-dependent corrections for a scalar loop.
Following this lead we will study two models with additional
scalars in this section.

%%%%%%%%%%%%%%%%%%%%%%%%%%%%%%%%%%%%%%%%%%%%%%%%%%%%%%%%%%%%
\subsection{Effective Lagrangian}
\label{sec:photons_eft}

In terms of the effective dimension-6 Lagrangian defined in
Eq.\eqref{eq:EFT} the effective Higgs--photon interaction is described
by the single operator $\ord_\gamma$. It is generated within the
Standard Model and by possible new physics particles at one loop.  The
corresponding Wilson coefficient is normalized such that the effective
$h\gamma\gamma$ interaction becomes 
\begin{align}
\lag \supset -\frac{1}{4v}\,\left( g_{h\gamma\gamma}^\text{SM} - \wilson_\gamma \, \frac{16\pi\,\aem v^2}{\Lambda^2} \right) \,F^{\mu\nu}\,F_{\mu\nu}\,H 
\label{eq:haaeff}
\end{align}
This coupling generates a modified $h\to \gamma\gamma$ decay width of 
\begin{align}
\Gamma(h\to \gamma\gamma) = \frac{m_H^3 G_F}{32\pi\,\sqrt{2}} \;
\left| g_{h\gamma\gamma}^\text{SM} - \wilson_\gamma \, \frac{16\pi\,\aem\,v^2}{\Lambda^2} \right|^2 \; .
\label{eq:haa-width}
\end{align}
As long as we are mostly interested in on-shell Higgs decays to
photons, there is little to learn from the kinematics of the two
photons. We therefore describe new physics effects as well as
differences between the full model and the dimension-6 approximation
in terms of
\begin{align}
\epsaa 
= \frac{\Gamma_{\gamma\gamma}}{\Gamma^\text{SM}_{\gamma\gamma}}-1 
= \dfrac{\left| g_{h\gamma\gamma}^\text{SM} - \wilson_\gamma \, \dfrac{16\pi\,\aem\,v^2}{\Lambda^2} \right|^2}{\left| g_{h\gamma\gamma}^\text{SM} \right|^2} - 1
\; .
 \label{eq:epsdef}
\end{align}
Notice that for decay processes that are loop-induced in the full
model, such as $h \to \gamma\gamma$ or $h\to \gamma Z$, there are no
additional contributions to $\epsaa$. Effects from mass pole residue
modifications or shifts in the SM input parameters, dubbed
\emph{residue} $\epsilon_R$ and \emph{parametric} $\epsilon_P$
corrections in Ref.\cite{hlm}, contribute to higher orders in the
effective Lagrangian. Similarly, the leading new physics contributions
do not modify the decay kinematics, hence there is no effect from the
phase space integration. All these aspects, combined with our
conservative choice of benchmark points mean that, unlike for
production-side contribution from effective
Lagrangians~\cite{Biekotter:2016ecg} we can linearize the new physics
effects in Eq.\eqref{eq:epsdef} without ruining the effective
Lagrangian approach altogether.

Since in the following we focus on additional scalars we can combine
Eq.\eqref{eq:haalag} and Eq.\eqref{eq:haaeff} to arrive at the general
structure of the matching condition
\begin{align}
\frac{\wilson_\gamma}{\Lambda^2} 
= \frac{C_S \,Q_S^2}{32 \pi^2} \, \frac{g_{hSS}}{v} \frac{A_S(\tau_S)}{m_S^2}
\approx  \frac{C_S \,Q_S^2}{192 \pi^2} \,\frac{g_{hSS}}{v} \frac{1}{m_S^2} \; ,
\label{eq:haamatch}
\end{align}
where, as usual, we will study the definition of the matching scale
$\Lambda$ and the treatment of terms suppressed by $v/\Lambda$ for
different new physics models.

%%%%%%%%%%%%%%%%%%%%%%%%%%%%%%%%%%%%%%%%%%%%%%%%%%%%%%%%%%%%
\subsection{Higgs doublet extension}
\label{sec:photons_2hdm}

The Higgs portal model discussed in Sec.~\ref{sec:oblique_portal} is
not well suited to study new physics effects in Higgs decays to
photons. The reason is that the additional state is not charged
and therefore does not contribute to the loop-induced
coupling. Therefore, we here instead consider an extended Higgs sector 
with a second doublet. It is convenient to work in the so-called
Higgs basis with the scalar potential~\cite{2hdm_basis}
\begin{align}
 V(H_1, H_2) 
&= Y_1\,H_1^\dagger\,H_1 
 + Y_2\,H_2^\dagger\,H_2 
 + Y_3\, \left( H_1^\dagger\,H_2 + \text{h.c.} \right) \notag \\
&+ \frac{Z_1}{2} \left( H_1^\dagger H_1\right)^2
 + \frac{Z_2}{2} \left(H_2^\dagger H_2 \right)^2 
 + Z_3\,\left( H_1^\dagger H_1\right) \left( H_2^\dagger H_2\right) 
 + Z_4\,\left( H_1^\dagger H_2\right) \left( H_2^\dagger H_1\right) \notag \\
&+ \left[ \frac{Z_5}{2} \left( H_1^\dagger\,H_2 \right)^2 
        + \left( Z_6\,H_1^\dagger\,H_1 + Z_7 H_2^\dagger\,H_2\right) H_1^\dagger\,H_2\, + \text{h.c.} \right] \; .
\label{eq:potential-higgs}
\end{align}
In this basis only the $H_1$ doublet develops a vev, $\langle H_1
\rangle = v$, while $\langle H_2 \rangle=0$.  In terms of the mass
eigenstates, the Higgs doublets can be expressed as
\begin{align}
H_1 = \left(\begin{array}{c} G^+ \\[2mm] \dfrac{v+h+iG^0}{\sqrt{2}} \end{array}\right)
\qqquad 
H_2 = \left(\begin{array}{c} \PHiggs^+ \\[2mm] \dfrac{H+i\Azero}{\sqrt{2}} \end{array}\right) \; ,
\label{eq:fields-higgsbasis}
\end{align}
and mapped back onto the generic basis $\{ \Phi_k \}$ through the rotation
\begin{align}
\begin{pmatrix} H_1  \\ H_2 \end{pmatrix}
= \begin{pmatrix} c_\beta & s_\beta \\ - s_\beta & c_\beta \end{pmatrix}
  \; \begin{pmatrix} \Phi_1  \\ \Phi_2 \end{pmatrix} \; ,
\label{eq:generic-to-higgs}
\end{align}
where now both of the doublets develop a non-zero vev $\braket{\Phi_k}
= v_k/\sqrt{2}$, with $v_1 = v s_\beta$, $v_2 = v c_\beta$.  In the
Higgs basis all quartic couplings are $SO(2)$--invariant. The
corresponding internal symmetries of the model can be thought of as
rotations in a Higgs flavor space. The relation
\begin{align}
s_{\beta-\alpha} \,c_{\beta-\alpha} = -\frac{Z_6\,v^2}{\mHHd - \mhd} 
\label{eq:limits}
\end{align}
neatly separates the decoupling limit $\mHH \gg v,
\mh$~\cite{Gunion:2002zf} from alignment without decoupling, $Z_6 \to
0$~\cite{Carena:2013ooa}. 

In general, there exist two sources of new physics contributions to
the decay rate $h\to \gamma\gamma$.  First, the SM-like Higgs
couplings to the $W$-boson and the heavy fermions are shifted through
the rotation of the Higgs mass eigenstates by an angle $\alpha$ and
the rotation of the vevs by an angle $\beta$.  Second, a charged Higgs
loop mediates the effective Higgs-photon coupling following
Eq.\eqref{eq:haalag}. In our analysis we will focus on the
alignment setup, removing the shifted SM-like couplings from our
analysis of the Higgs--photon coupling. The only remaining effect then
is the charged Higgs loop contribution. For the two-Higgs-doublet model the loop
contribution from the charged Higgs is mediated by the triple scalar
coupling
\begin{align}
g_{h H^+H^-} 
= \frac{1}{v}\,
  \left(m_h^2+2m^2_{H^{\pm}} - \frac{2 m^2_{12}}{s_\beta c_\beta}\right) 
= v Z_3 
\qquad \text{with} \quad m^2_{H^{\pm}} = Y_2 + \frac{Z_3 v^2}{2} \; ,
\label{eq:hhphm}
\end{align}
up to corrections of $\ord (c^2_{\beta-\alpha})$.  The dimension-two
coefficient $Y_2$ generates the heavy doublet mass scale in the
gauge-symmetric phase. In the more familiar basis parameters~\cite{2hdm_review} it is
given by $Y_2 = m^2_{11}\,c^2_\beta +m^2_{22}\,s^2_\beta + m_{12}^2\,
s_{2\beta}$. It also gives the default matching scale in the unbroken
phase, $\Lambda^2 = Y_2$.\bigskip

As always, we illustrate different matching schemes, attempting to
systematically improve the agreement between full model and
dimension-6 Lagrangian:

\begin{itemize}
\item SP1: \textit{default matching,} where the full model is matched
  to the dimension-6 effective Lagrangian in the unbroken
  phase. Following the general structure of Eq.\eqref{eq:haamatch} and
  assuming the matching scale $\Lambda^2 = Y_2$, we find the relevant
  Wilson coefficient
\begin{align}
\frac{\wilson_\gamma}{\Lambda^2} = \frac{Z_3}{192\pi^2}\,\frac{1}{Y_2} \; .
\label{eq:2hdm-default}
\end{align} 

\item BP1$v$: \textit{$v$-improved broken-phase matching,} where we
  obtain $\wilson_\gamma$ from the (derivative of the) 1PI photon
  Greens function in the full model and the effective Lagrangian
  setups, expanding both sides of the identity to
  $\mathcal{O}(v^2/\Lambda^2)$, and identifying the matching scale $\Lambda$
  with the charged Higgs mass 
\begin{align}
& \frac{d\Pi_{\gamma\gamma}}{dp^2}\Big{\lvert}_{p^2=0} = \frac{d\Pi^\text{EFT}_{\gamma\gamma}}{dp^2}\Big{\lvert}_{p^2=0}
= \frac{8\,\swd\,m_W^2}{\Lambda^2} \wilson_\gamma \quad 
\Rightarrow\qquad \frac{\wilson_\gamma}{\Lambda^2} = \frac{Z_3}{192\pi^2} \,\frac{1}{m^2_{H^{\pm}}} \; .
\label{eq:kgamma-2hdm}
\end{align}

\end{itemize}
\bigskip

%-----------------------------------------------------
\begin{table}[b!]
 \begin{tabular}{c|rrrrrrr|ccc} \hline
   &  $m_H$ & $m_A$ & $m_{H^{\pm}}$ & $\tan\beta$ & $m^2_{12}$ & $\sqrt{Y_2}$ & $Z_3$ & \multicolumn{3}{c}{$\epsaa$} \\
   & & & & & & & &  full model & SP1  & BP1$v$  \\ \hline
D1 & 350 & 350 & 350 &  2   & $4.9\times 10^{4}$ & 338.6 & 0.27 &  $6.62(60)\times 10^{-3}$ & $6.95\times 10^{-3}$ & $6.50\times 10^{-3}$ \\
D2 & 350 & 350 & 350 &  1.5 & $2.8\times 10^{4}$ & 231.2 & 2.36 &  $5.85(76)\times 10^{-2}$ & $6.56\times 10^{-5}$ & $5.75\times 10^{-2}$ \\ \hline
 \end{tabular}
\caption{Benchmark points and predictions for $\epsaa$, as defined in
  Eq.\eqref{eq:epsdef}, in the 2HDM model and its different matching
  setups. All masses are given in GeV. For the full model, the digit
  in brackets accounts for the square of the charged Higgs
  contribution.}
 \label{tab:2hdm-results}
\end{table}
%-----------------------------------------------------

As above, we compute the deviation in the Higgs-photon coupling
$\epsaa$, defined in Eq.\eqref{eq:epsdef}, in the full model and in
the different matching setups.  Two benchmark points defined in
Tab.~\ref{tab:2hdm-results} represent two complementary regimes: The
first point D1 features a weakly coupled scenario, where the physical
heavy Higgs masses are driven by the doublet mass $Y_2$. The second
point D2 is strongly coupled, and a sizable fraction of the heavy
Higgs mass is generated by non-decoupling contributions proportional
to $v$.  Both scenarios satisfy all theoretical and experimental
constraints on the model, in particular the charged Higgs mass limits
from direct searches~\cite{Abdallah:2003wd} and flavor
observables~\cite{flavor}. Additionally, in the
alignment limit the lightest CP-even mass eigenstate exactly mimics
the properties of the SM Higgs, and therefore is in excellent
agreement with the LHC data.

The alignment condition fixes $\alpha = \beta-\pi/2$.  In this limit,
without any mixing between the two doublets, it makes no difference
whether $\wilson_\gamma$ is obtained via explicit matching or by
integrating out the heavy doublet in the unbroken phase with standard
functional methods~\cite{Gorbahn:2015gxa}.  

Since the Higgs couplings to fermions play no role here, we do not
need to choose a specific setup for the Yukawa couplings. Also, with
no loss of generality, we may assume all heavy Higgs companions to be
mass-degenerate.\bigskip

We show results for the full model and its two different matching
schemes in Tab.~\ref{tab:2hdm-results}. First, for the full model we
see that taking into account the squared term $\propto
\wilson_\gamma^2$ in Eq.\eqref{eq:epsdef} has no measurable effect,
because the charged Higgs effects in general hardly reach the per-cent
level. Note that if we attempt to define a benchmark point with
order-one deviations from the Standard Model, this picture will of
course change, and we would have to adapt our
approach~\cite{Biekotter:2016ecg}.

For the weakly interacting benchmark point D1, the full model
prediction is quite accurately reproduced by the effective Lagrangian
in either of the two matching schemes.  For the strongly interacting
point D2 the charged Higgs contributions are driven by sizable
$v$-mediated couplings.  The deviations in $\epsaa$ are one order of
magnitude larger than in D1, and the squared terms in $\wilson_\gamma$
gain a little more relevance.  The large values of $Z_3$, along with
the sizeable split between the default matching scale $\sqrt{Y_2}$ and
the charged Higgs mass, explain the sizeable difference between the
full model and SP1 matching. On the other hand, broken phase matching,
BP1$v$, leads to significant improvement over SP1.

%%%%%%%%%%%%%%%%%%%%%%%%%%%%%%%%%%%%%%%%%%%%%%%%%%%%%%%%%%%%
\subsection{Scalar top partners}
\label{sec:photons_partners}

As a second example, we consider the toy model extending the Standard
Model by a set of scalar top partners, introduced in
Sec.~\ref{sec:oblique_partners}. In the spirit of minimal flavor
violation, we assume that only the scalar top partners, but not the
bottom partner, have sizeable Higgs couplings,
\begin{align}
g_{h\gamma\gamma} = -\frac{\aem}{\pi}\,
 \left[ \sum_{f = t,b,\tau}\, C_f \,Q_f^2 A_f(\tau_f) 
       + A_W(\tau_W) 
       + \sum_{\tilde{t}}
       C_{\tilde{t}}\,Q^2_{\tilde{t}}\,\frac{g_{h\tilde{t}\,\tilde{t}}\,v}{2m^2_{\tilde{t}}}\, A_S(\tau_{\tilde{t}}) \right] \; .
 \label{eq:haalag-partners}
\end{align}
The heavy top partners couple to the Higgs boson through the
off-diagonal entries in their mass matrix,
\begin{align}
\frac{g_{h\stone\stone}}{v} = \kLL\,\ctd + \kRR\,\std + \frac{\kLR}{\sqrt{2} v} \sdt 
\qqquad \text{and} \qqquad 
\frac{g_{h\sttwo\sttwo}}{v} = \kLL\,\std + \kRR\,\ctd - \frac{\kLR}{\sqrt{2} v} \sdt  \; .
 \label{eq:hsqsq}
\end{align}
Along the lines of Eq.\eqref{eq:haaeff} we can relate the
Higgs--photon couplings in the dimension-6 Lagrangian to the full top
partner model as 
\begin{align}
\frac{\wilson_\gamma}{\Lambda^2} 
= \frac{1}{24 \pi^2 v}\,\left[\frac{g_{h\stone\stone}}{m^2_{\stone}}\,A_S(\tau_1)
  + \frac{g_{h\sttwo\sttwo}}{m^2_{\sttwo}}\,A_S(\tau_2)\right] \; .
 \label{eq:partners-cgammafull}
\end{align}
To see how accurately the full model prediction for the Higgs-photon
coupling $\epsaa$ is approximated by effective Lagrangian we consider
the same matching setups as in Sec.~\ref{sec:oblique_partners}:

%-----------------------------------------------------
\begin{figure}[t]
\includegraphics[width=0.4\textwidth]{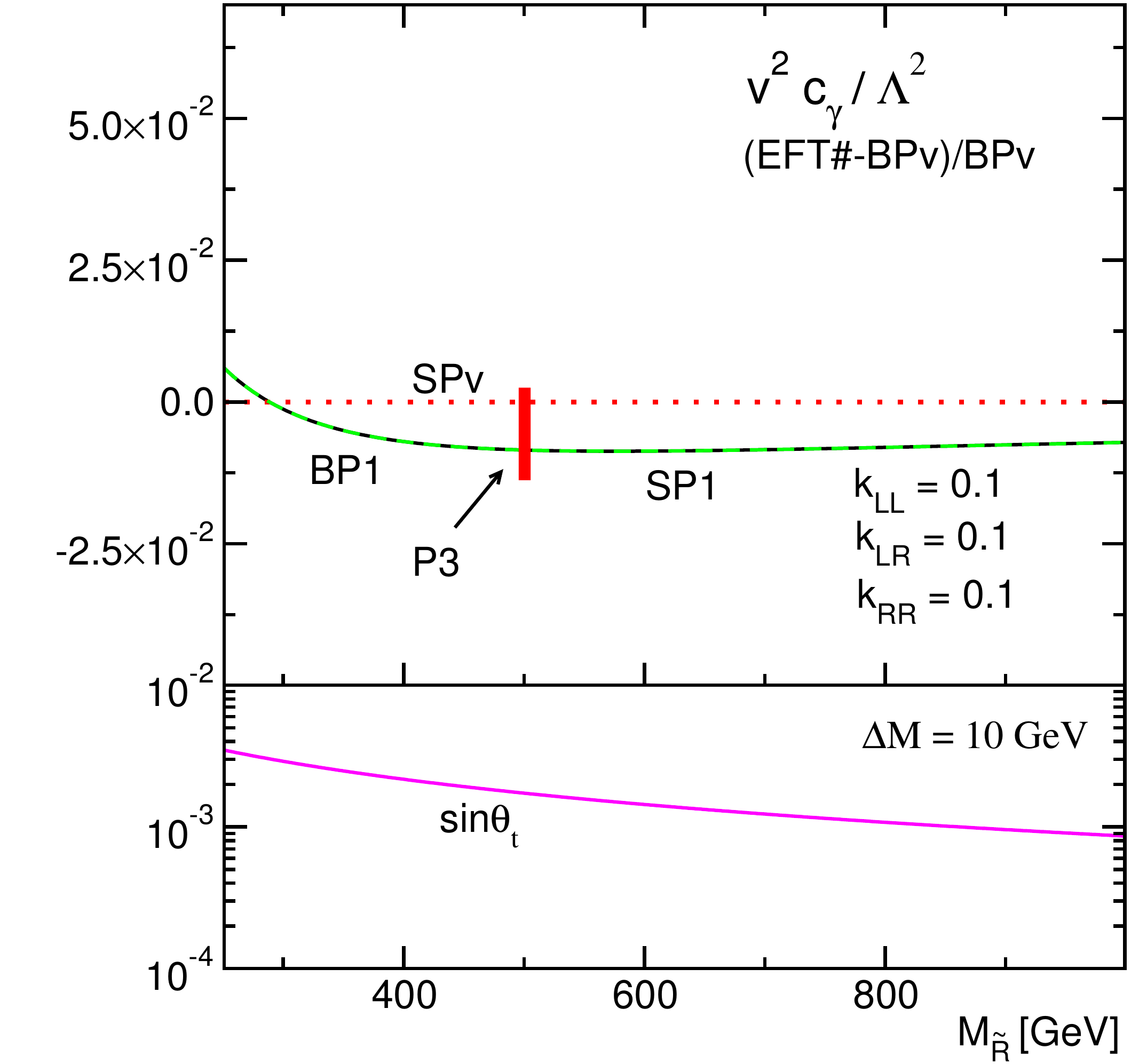} 
\hspace*{0.08\textwidth}
\includegraphics[width=0.4\textwidth]{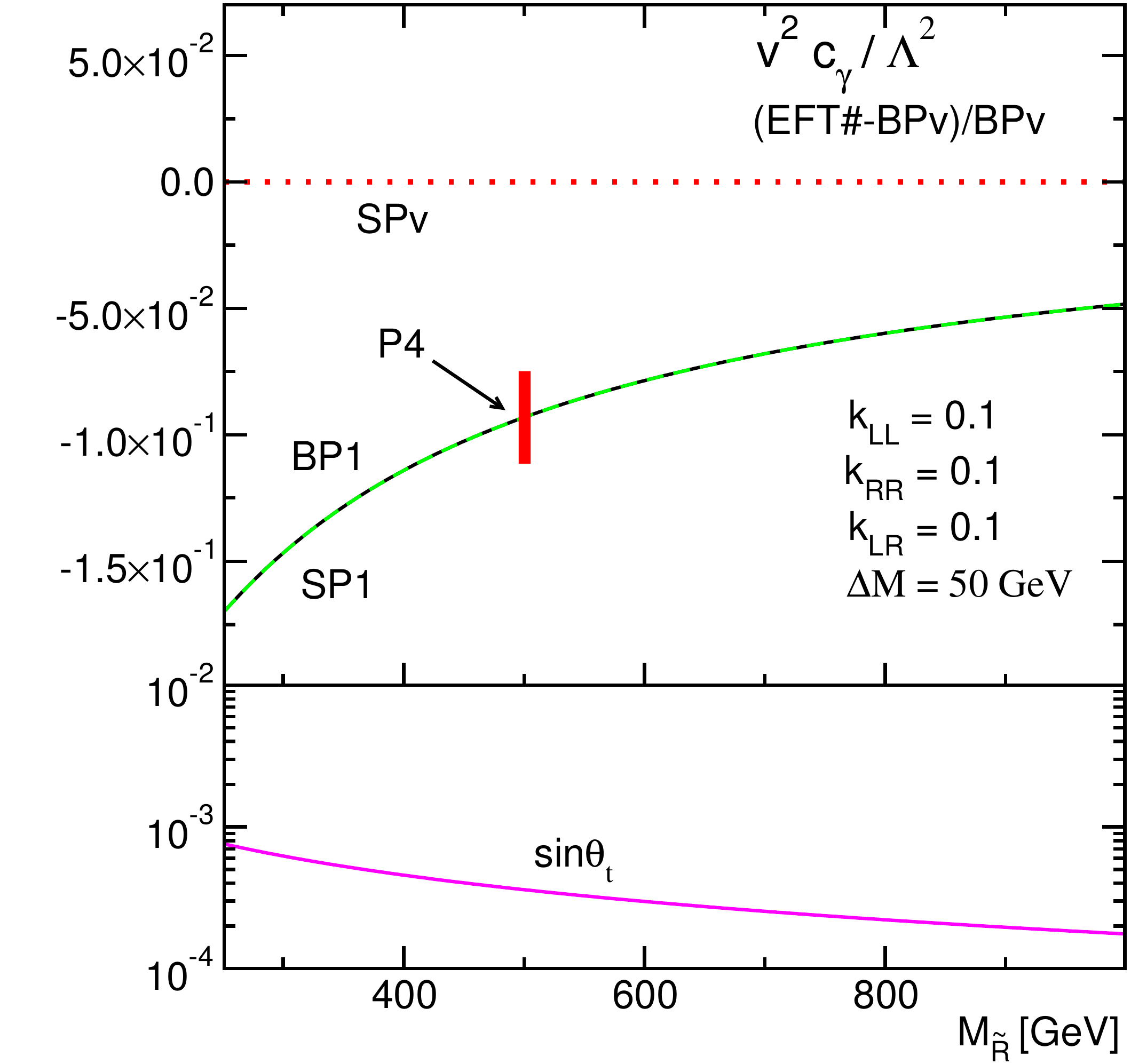} 
\caption{Wilson coefficient $\wilson_\gamma$ as a function of the
  heavy partner mass $\MRight$.  The different curves show the
  relative deviation of the different effective Lagrangian
  predictions relative to the most accurate setup BP$v$.  We consider
  two mass splittings $\Delta M \equiv \MRight- \MLeft = 10$~GeV
  (left) and 50~GeV (right). The red marks indicate the benchmark
  points P3 and P4 from Tab.~\ref{tab:partners-benchmarks}. The
  decoupling behavior of the stop mixing angle is shown in the lower
  sub-panels.}
\label{fig:wilson-partners-decoupling}
\end{figure}
%-----------------------------------------------------

\begin{itemize}
\item SP1: \textit{default matching,} in which the full model is
  matched to the dimension-6 effective Lagrangian in the unbroken
  phase at $\Lambda = M$. We assume a common
  heavy spectrum $\MLeft = \MRight \equiv
  M$~\cite{hlm},
 \begin{align}
\frac{\wilson_\gamma}{\Lambda^2} 
= \frac{1}{144\,\pi^2\,M^2}\left[ \kLL + \kRR - \frac{\kLR^2}{M^2} \right] \; .
 \label{eq:kgamma-eft1}
 \end{align}
\item SP2: \textit{non-degenerate masses,} where we work again in the
  unbroken phase, but integrate out non-degenerate heavy fields with
  $\MLeft \neq \MRight$ separately~\cite{Drozd:2015kva,Drozd:2015rsp}
 \begin{align}
\frac{\wilson_\gamma}{\Lambda^2} 
= \frac{1}{144\,\pi^2}
\left[ \frac{\kLL}{M^2_{\tilde{Q}_L}} + \frac{\kRR}{M^2_{\tilde{T}_R}}
-\frac{\kLR^2}{M^2_{\tilde{Q}_L}\,M^2_{\tilde{T}_R}} \right] \; .
\label{eq:kgamma-eft2}
 \end{align}

\item SP$v$: \textit{$v$-improved matching,} which starting from the
  above result is defined through the replacements $\MLeft \to
  \mst{1}$, $\MRight \to \mst{2}$, $\kLL \to \tilde{\kappa}_{LL}$, and
  $\kRR \to \tilde{\kappa}_{RR}$
 \begin{align}
\frac{\wilson_\gamma}{\Lambda^2} 
= \frac{1}{144\,\pi^2}
\left[ \frac{ \ctd\kLL + \std\kRR}{\mst{1}^2} + \frac{\std\kLL + \ctd\kRR}{\mst{2}^2}
-\frac{\kLR^2}{\mst{1}^2 \mst{2}^2} \right]  \; .
\label{eq:kgamma-eft3}
 \end{align}

\item BP1: \textit{broken-phase matching,} in which case the Wilson
  coefficients are derived through explicit matching in the broken
  phase. For a single heavy mass scale $M$ we find
 \begin{align}
\frac{\wilson_\gamma}{\Lambda^2} 
  = \frac{1}{144\,\pi^2\,M^2}
    \left[ \kLL + \kRR - \frac{\kLR^2\,\sdt^2}{M^2} \right] \; . 
\label{eq:kgamma-eft4} 
 \end{align}

 \item BP$v$: \textit{$v$-improved  broken-phase matching,} where the
   different heavy scales in
 \begin{align}
\frac{\wilson_\gamma}{\Lambda^2} 
&= \frac{1}{144\,\pi^2}
   \left[ \frac{\ctd\kLL + \std\kRR}{\tilde{M}_L^2} - \frac{\kLR^2\,\sdt^2}{2\tilde{M}_L^4}
        + \frac{\std\kLL + \ctd\kRR}{\tilde{M}_R^2} - \frac{\kLR^2\,\sdt^2}{2\tilde{M}_R^4} 
   \right]           \notag \\
  \label{eq:kgamma-eft5}
 \end{align}
  are now given by
 \begin{align}
  \tilde{M}_L^2 &= \MLeft^2\ctd + \MRight^2\std \to \mst{1}^2\ctd + \mst{2}^2\std \; \notag \\
  \tilde{M}_R^2 &= \MLeft^2\std + \MRight^2\ctd \to \mst{1}\std + \mst{2}^2\ctd \; .
 \label{eq:kgamma-eft6}
 \end{align}

\end{itemize}

%-----------------------------------------------------
\begin{figure}[t]
\includegraphics[width=0.4\textwidth]{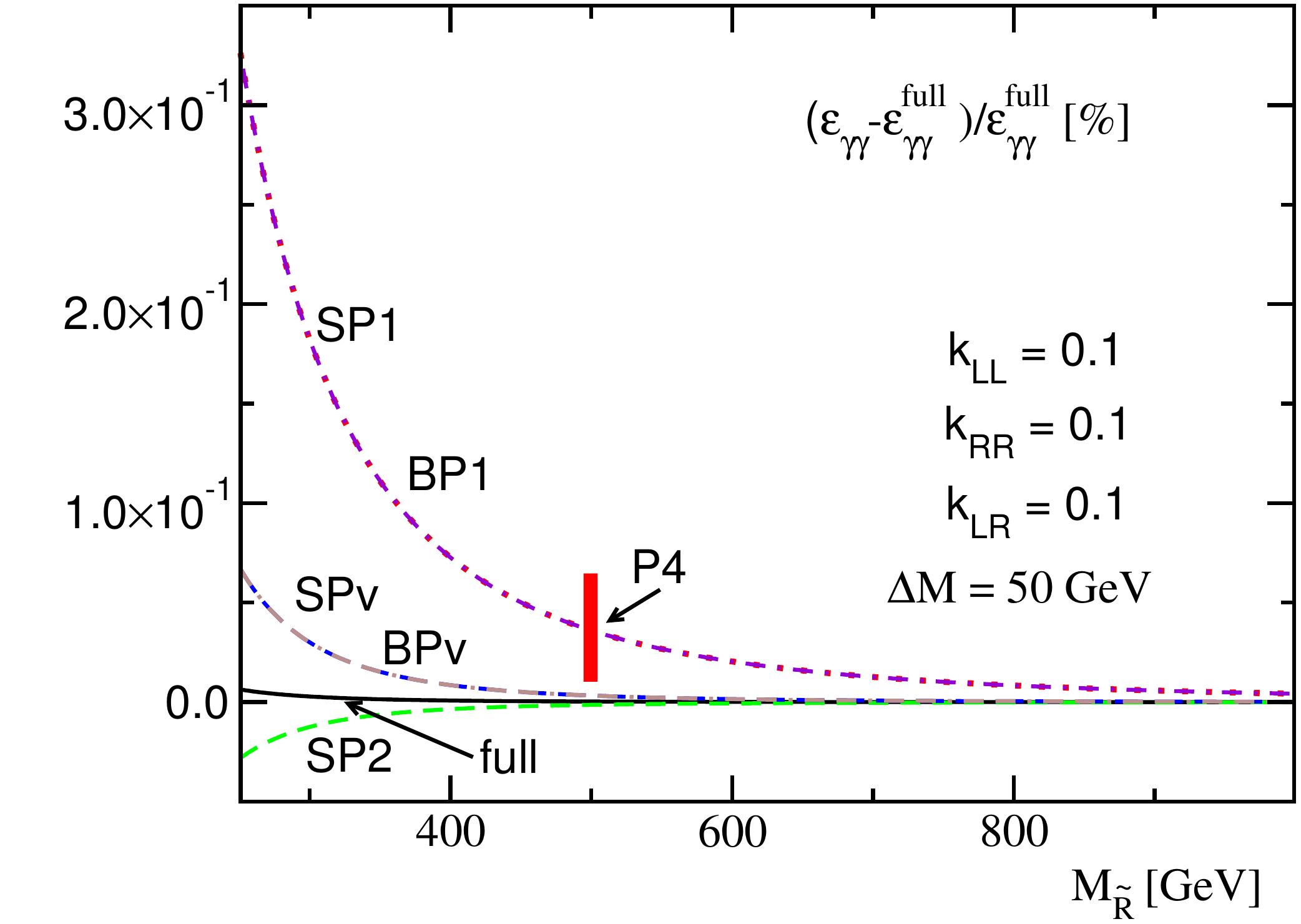} 
\hspace*{0.08\textwidth}
\includegraphics[width=0.4\textwidth]{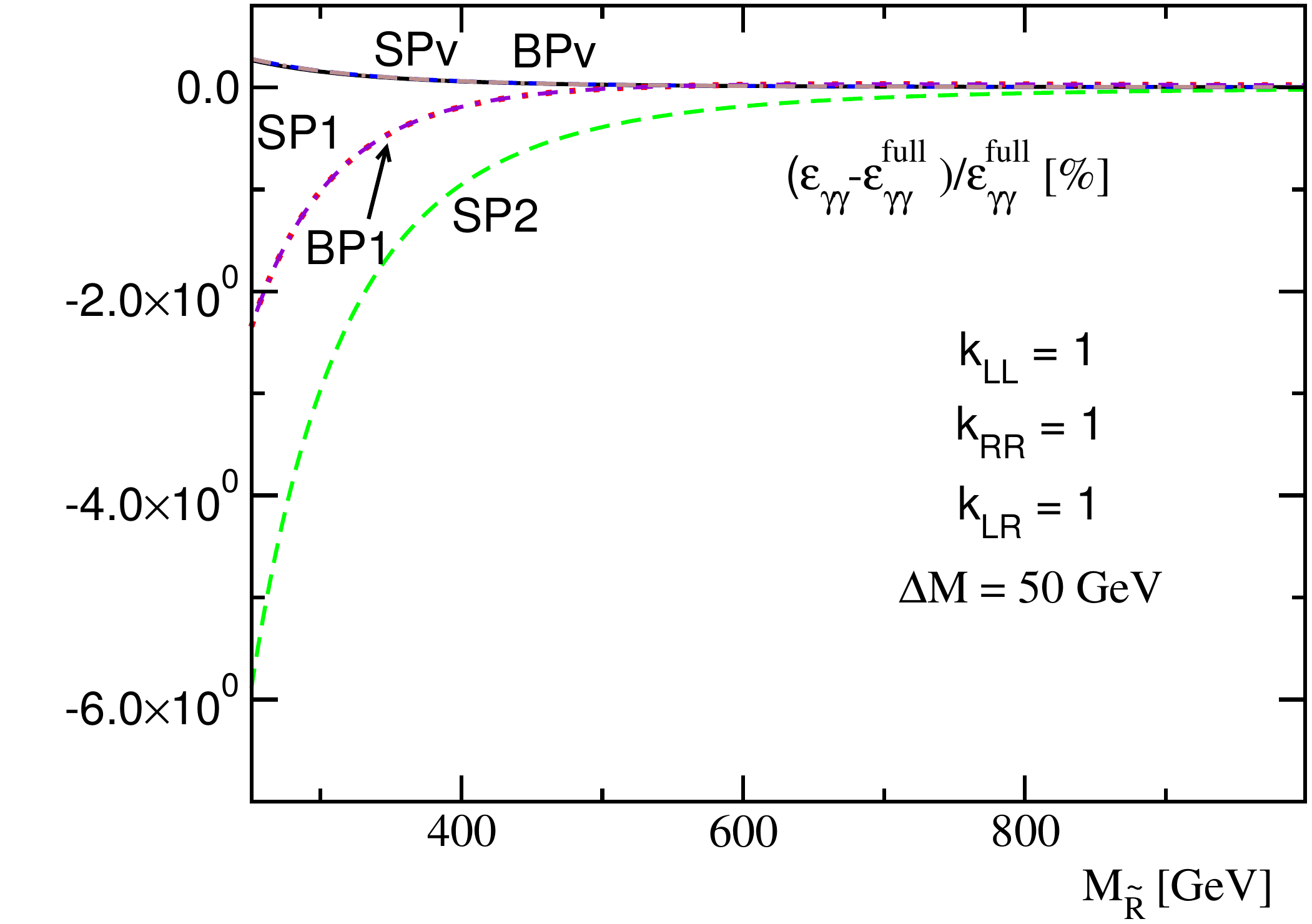} 
\caption{Predictions for $\epsaa$, as defined in Eq.\eqref{eq:epsdef},
  as a function of the heavy partner mass $\MRight$. The different
  curves show the relative deviation between the full model
  and the effective theory predictions for the different setups.  The
  partner couplings to the Higgs bosons at fixed to weak
  $\kLL,\kLR,\kRR = 0.1$ (left) and moderate $\kLL,\kLR,\kRR = 1$
  values (right). The red mark indicates the benchmark
  points P4 from Tab.~\ref{tab:partners-benchmarks}.}
\label{fig:epsaa-partners-decoupling}
\end{figure}
%-----------------------------------------------------

%-----------------------------------------------------
\begin{table}[b!]
 \begin{tabular}{l|rrrrrr} \hline
 $\epsaa$ & full model & SP1 & SP2 & SP$v$ & BP1 & BP$v$ \\ \hline   
P1 & $ 0.565(6)\times 10^{-2}$ & $ 0.538\times 10^{-2}$ & $ 0.538\times 10^{-2}$ & $ 0.560\times 10^{-2}$ & $ 0.538\times 10^{-2}$ & $ 0.560\times 10^{-2}$\\
P2 & $-0.354(1)\times 10^{-1}$ & $-0.466\times 10^{-1}$ & $-0.466\times 10^{-1}$ & $-0.349\times 10^{-1}$ & $-0.469\times 10^{-1}$ & $-0.356\times 10^{-1}$\\
P3 & $-0.324(4)\times 10^{-2}$ & $-0.319\times 10^{-2}$ & $-0.325\times 10^{-2}$ & $-0.322\times 10^{-2}$ & $-0.319\times 10^{-2}$ & $-0.322\times 10^{-2}$\\
P4 & $-0.355(5)\times 10^{-2}$ & $-0.319\times 10^{-2}$ & $-0.356\times 10^{-2}$ & $-0.352\times 10^{-2}$ & $-0.319\times 10^{-2}$ & $-0.352\times 10^{-2}$\\\hline
  \end{tabular}
 \caption{Predictions for $\epsaa$ in the scalar partner model and the
   different effective Lagrangian setups described in the text. The
   benchmark points are defined in
   Tab.~\ref{tab:partners-benchmarks}. The digit in brackets accounts
   for the squared of the scalar partner loops.}
 \label{tab:spartners-epsaa}
\end{table}
%-----------------------------------------------------

For these different matching setups we compute the modifications to
the Higgs-photons coupling $\epsaa$, as defined in
Eq.\eqref{eq:epsdef}. We use the same benchmarks as in
Sec.~\ref{sec:oblique_partners} for the results shown in
Tab.~\ref{tab:spartners-epsaa}.  The digit in parentheses indicates
the change when we add the square of the scalar partner loops. The
numerical results are similar to those of the oblique parameters in
Sec.~\ref{sec:oblique_partners}.  The very mild offsets between the
simple $v$-improved setup SP$v$ and the full broken phase matching
scheme BP$v$ can be attributed to the $\theta_{\tilde{t}}$-suppressed
contributions in BP$v$.  On the other hand, in contrast to the oblique
parameters, we find hardly any effect from a non-degenerate
spectrum with large mixing. This is related to the diagonal structure of the
electromagnetic coupling, which implies that, unlike
$c_{\text{W,B,T}}$, the one-loop contributions to $\Pi_{\gamma\gamma}$
do not feature a simultaneous exchange of different mass states. 
Our effective Lagrangian result for $c_\gamma$ therefore agrees very with the
full model, even for the strongly-coupled scenario P2.\bigskip

In Fig.~\ref{fig:epsaa-partners-decoupling} we study the decoupling
behavior of $\epsaa$ as a function of the heavy partner mass
$\MRight$. We consider a weakly-coupled scenario with $\kLL,\kLR,\kRR
= 0.1$, as in benchmarks P3 and P4, and compare it to the case of
$\kappa \sim \ord(1)$.  We find that the dimension-6 approximation
gives an excellent approximation to the full model for $M \gtrsim
400$~GeV for weak couplings and $M \gtrsim 600$~GeV for strong
couplings.  A comparably more dramatic breakdown of the effective
Lagrangian appears for $M \lesssim 400$~GeV in the strongly-coupled
case. Large couplings combined with a small scale separation render
the default matching approach inadequate, whereas the $v$-improved
matching agrees with the full model within less than 1\% down to $M
\approx 250$~GeV. In contrast to the oblique parameters, no delayed
decoupling is encountered even for large $v$-induced mass
splitting. As mentioned above, this is due to the fact that the two
scalar top partner eigenstates do not mix in the $h\gamma\gamma$
loops.

Some other characteristic trends already encountered in the
electroweak precision analysis are again visible in $\epsaa$: first,
the sign of the deviation flips between the default EFT truncation and
the $v$-improved matching, where the latter reproduces the full model
predictions much more accurately. Second, the broken-phase corrections
from explicit matching are again negligible since there are no mixed
heavy-light loops in the scalar top partner model. For the weakly
coupled scenario in the left panel of
Fig.~\ref{fig:epsaa-partners-decoupling}, we observe unexpectedly good
agreement between the SP2 matching, without $v$-improvement, and the
full model. However, this turns out to be simply a numerical
coincidence, as can be seen by inspecting the relatively poor
performance of SP2 in the scenario in
Fig.~\ref{fig:epsaa-partners-decoupling}~(right).

%%%%%%%%%%%%%%%%%%%%%%%%%%%%%%%%%%%%%%%%%%%%%%%%%%%%%%%%%%%%
\section{Summary}
\label{sec:summary}

To justify using an effective Lagrangian, for example truncated at
dimension six, we need to either show that higher-dimensional
contributions are negligible, or that the effective Lagrangian
reproduces the features of classes of complete models. In the second
case, the appropriate matching procedure can play a key role, in
particular if we integrate out particles right around the scale of
electroweak symmetry breaking, \ie at a scale where the structure of
the Lagrangian changes significantly. 

At tree level it is known that taking into account terms of the order
$v^2/\Lambda^2$ in the definition of the matching scale and in the
matching condition can make a sizeable
difference~\cite{Brehmer:2015rna}. In this paper we have
systematically studied possible improvements in the matching procedure
at the one-loop level, considering the oblique electroweak parameters
$S$ and $T$, as well as the Higgs decay width to photons. For
extended scalar sectors we have confirmed three ways to systematically
improve the matching procedure of a dimension-6 Lagrangian with
linearly realized electroweak symmetry breaking:
\begin{enumerate}
\item $v$-improving the matching scale and matching condition by
  expressing them in terms of the (lightest) integrated-out particle
  mass and mixing angles
  will improve the agreement with the full model, both at tree level
  and at loop level;
\item determining the matching condition based on Greens functions in
  the broken phase and including the appropriate $v$-suppressed terms
  can lead to a systematic improvement, if combined with $v$-improved
  matching;
\item properly taking into account several new physics scales, if
  present, to avoid issues with mass splittings induced in the unbroken
  as well as in the broken phase.
\end{enumerate}
Simple and convenient matching schemes based on leading
logarithms~\cite{llew,llhiggs}, in contrast, are not useful for any
kind of precision physics. Altogether we have introduced an
appropriate matching procedure around the scale of electroweak
symmetry breaking, systematically including $v$-induced effects. We
note that it is always possible to find better agreement between full
models and the effective with the help of $v$-improvement, but the
appropriate form of the $v$-suppressed terms depends on the model and
the number of scales involved.  In particular for the LHC the freedom
to optimize the matching procedure will be the key to defining a
usable effective Lagrangian approach.\bigskip

%%%%%%%%%%%%%%%%%%%%%%%%%%%%%%%%%%%%%%%%%%%%%%%%%%%%%%%%%%%%
\begin{center} \textbf{Acknowledgments} \end{center}

First of all we would like to thank Juan Gonz\'alez-Fraile for his
contributions during an early phase of the project. TP, AF, and DLV
would like to thank the MITP, KITPC, and Institut f\"ur Theoretische
Physik at Karlsruhe for their hospitality while this paper was
finished. DLV is funded by the F.R.S.-FNRS \emph{Fonds de la Recherche
  Scientifique} (Belgium). The work of AF is funded in part by the
U.S.\ National Science Foundation under grant PHY-1519175. DLV also
wishes to warmly thank Cen Zhang for more than enlightening
discussions.

%%%%%%%%%%%%%%%%%%%%%%%%%%%%%%%%%%%%%%%%%%%%%%%%%%%%%%%%%%%%

\end{document}